\documentclass[useAMS,usenatbib]{mn2e}
\usepackage{graphics,epsfig,psfig}
\usepackage[normalem]{ulem}
\usepackage[dvipsnames]{color}
\usepackage{xcolor}
\usepackage{amsmath, amssymb}
\usepackage[]{inputenc,amssymb}
\usepackage{booktabs,caption}

\def \be{\begin{equation}}
\def \ee{\end{equation}}
\def \bea{\begin{eqnarray}}
\def \eea{\end{eqnarray}}
\def \etal{{et al.}}

\definecolor{webgreen}{rgb}{0,.5,0}
\definecolor{webbrown}{rgb}{.6,0,0}

\usepackage[%
    colorlinks = true,%
    linkcolor = blue,%
    urlcolor  = blue,%
    citecolor = webgreen,%
    anchorcolor = blue]{hyperref}
\newcommand{\ufhref}[3][blue]{\href{#2}{\color{#1}{#3}}}%

\title[Effects of radiation on superbubbles]{How radiation affects superbubbles : Through momentum injection in early phase and photo-heating thereafter}
\author[Gupta, Nath, Sharma \& Shchekinov]
{Siddhartha Gupta$^{1,2}$ \thanks{E-mail : siddhartha@rri.res.in}, Biman B. Nath$^1$, Prateek Sharma$^2$, Yuri Shchekinov$^3$\\
$^1$Raman Research Institute, Sadashiva Nagar, Bangalore 560080, India\\
$^2$Joint Astronomy Programme and Department of Physics, Indian Institute of Science, Bangalore 560012, India\\
$^3$P. N. Lebedev Physical Institute, 53 Leninskiy Prospekt, 119991 Moscow, Russia
}
\begin{document}
\maketitle
\label{firstpage}
\begin{abstract}
\noindent Energetic winds and radiation from massive star clusters push the surrounding gas and blow superbubbles in the interstellar medium (ISM). Using 1-D hydrodynamic simulations, we study the role of radiation in the dynamics of superbubbles driven by a young star cluster of mass $10^{6}$ M$_{\odot}$. We have considered a realistic time evolution of the mechanical power as well as radiation power of the star cluster, and detailed heating and cooling processes. We find that the ratio of the radiation pressure on the shell (shocked ISM) to the thermal pressure ($\sim10^{7}$ K) of the shocked wind region is almost independent of the ambient density, and it is greater than unity before $\lesssim 1$ Myr. We explore the parameter space of density and dust opacity of the ambient medium, and find that the size of the hot gas ($\sim$ 10$^{7}$ K) cavity is insensitive to the dust opacity ($\sigma_{d}\approx(0.1-1.5)\times 10^{-21}$ cm$^{2}$), but the structure of the photoionized ($\sim10^4$ K) gas depends on it. Most of the radiative losses occur at $\sim10^{4}$ K, with sub-dominant losses at $\lesssim 10^3$ K and $\sim10^{6}-10^{8}$ K. The superbubbles can retain as high as $\sim 10\%$ of its input energy, for an ambient density of $10^{3}\,m{\rm_{H}\,cm^{-3}}$. We discuss the role of ionization parameter and recombination-averaged density in understanding the dominant feedback mechanism. Finally, we compare our results with the observations of 30 Doradus.
\end{abstract}
\section{Introduction}
The study of interactions between stars and their surrounding medium is crucial in understanding the evolution of galaxies. The formation of H{\scriptsize {II}} regions (\citealt{Stromgren1939}) and the expansion of giant gas shells (\citealt{Castor1975}) are manifestations of these interactions. These feedback processes affect subsequent star formation as well as control the chemical enrichment of galaxies (\citealt{Hopkins2011}; \citealt{Federrath2015}; \citealt{Skinner2015}; \citealt{Dale2015}).

An important aspect of this interaction is mechanical feedback, that arises from the mechanical energy deposited by the stars in the form of stellar winds and supernovae (\citealt{Taylor1950}; \citealt{Parker1965}). The ionizing radiation of massive stars is so strong that it can remove the surrounding gas before the supernova occurs (\citealt{Geen2015}). For a young star cluster, the bolometric luminosity ($L_{bol}$) is $\sim 100$ times larger than the mechanical luminosity ($L_{w}$, c.f. Figure \ref{fig:outSB99}), and  the momentum deposition rate due to radiation $\sim L_{bol}/c$ is almost comparable to the momentum deposition rate due to mechanical energy $\sim L_w/v_w$ ($v_w$ is the cluster wind velocity). However, previous authors have either considered the effect of the winds (\citealt{MacLow1988}; \citealt{Rogers2013}) or that of the radiation (\citealt{Dale2013}; \citealt{Sales2014}). The combined effect of radiation and winds have not been studied in detail in a simple set up.

A useful way to characterize the importance of wind and radiation is to compare thermal and radiation pressures in the bubble. The supersonic winds interact with the ambient medium, produce a shock wave and sweep up the surrounding matter into a thin shell (shocked ISM, hereafter `shell'). During this process, the wind loses kinetic energy, gets thermalized and thereby a high-pressure zone (shocked wind region, hereafter `SW') is formed. The thermal pressure in the SW region ($P_{sw}$) is expected to scale as $\rho_{amb}\,(L_w/\rho_{amb}t^2)^{2/5} \propto R_{cd}^{-4/3}$ ($R_{cd}$ is the location of the contact discontinuity between the shocked ISM and SW, $\rho_{amb}$ is density of the ambient medium); (\citealt{Weaver1977}). The radiation pressure ($P_{rad}$) can be estimated as $L_{bol}/(4\pi\,R_{cd}^2\,c)$, which falls ($\propto R_{cd}^{-2}$) faster than $P_{sw}$ as the bubble expands. Therefore, the radiation pressure is believed to be important at early times . In a dense ISM, the radiation pressure can be more important because radiative cooling removes a large fraction of mechanical energy (e.g., \citealt{Yadav2016}).

Some authors have highlighted the role of radiation pressure for massive star clusters (\citealt{Krumholz2009}; \citealt{Murray2011}). \citealt{Lopez2011} observed 30 Doradus and concluded that the radiation pressure have a significant role on its dynamics. \citealt{Pellegrini2011} also analysed the same object and by modelling different line ratios found that radiation pressure is weak compared to thermal pressure of the X-ray emitting gas. \citealt{Yeh2012} measured ionization parameter of local starburst galaxies and concluded that the radiation pressure seems to be important.

For a better understanding of the relative role of mechanical and radiation feedback, \citealt{Silich2013} (hereafter ST13) revisited the standard model of interstellar bubbles (ISB); (\citealt{Weaver1977}). They found that in a dense medium the SW region cools rapidly, thereby draining its thermal energy and making it disappear. They concluded that the radiation pressure is important only in such a case when there is no SW region. In their example of a superbubble being driven by a cluster of mass $10^6$ M$_\odot$, in an ambient medium of particle density $10^{3}$ cm$^{-3}$, radiation pressure becomes important during a short interval of $\approx$ $1.6\hbox{--}2$ Myr. However, their estimates of the cooling time scale of the SW region are based on the adiabatic calculations of \citealt{MacLow1988}, and it is not clear if they hold in the presence of radiative cooling in the dense shell.

In this work, using 1-D hydrodynamic simulations, we present more realistic calculations and discuss the importance of thermal conduction, radiative cooling, heating, and radiation pressure. We find that, in dense media ($\rho_{amb}\gtrsim 10^{2}\,m\rm{_{H}\, cm^{-3}}$), despite the rapid cooling of the shell, the SW region does not disappear. We also find that the ratio of the radiation pressure to thermal pressure ratio is greater than unity before $\lesssim 1$ Myr. This paper improves the understanding of ISBs in the dense medium.

The plan of this paper is as follows. We start with a comparative study of the constant luminosity model and a more realistic time-dependent luminosity model. In the case of time-dependent luminosity, the radiation power and mechanical power of a typical star cluster are obtained using Starburst99 (\citealt{Leitherer99}), the details of which are discussed in section \ref{sec:setupSb99}. In section \ref{sec:analytic}, we discuss the analytical model of an interstellar bubble. In section \ref{sec:simsetup}, we describe our simulation set-up. The results of simulations are discussed in section \ref{sec:simresult}. In section \ref{sec:modlobscomp} we explore the parameter space, compare our results with other theoretical models and with observations. Finally, in section \ref{sec:summary}, we summarise the main results of this paper.
\section{Set-up \& output of Starburst99}
\label{sec:setupSb99}
We use Starburst99 v7.0.1 (\citealt{Leitherer99}) to generate the mechanical and radiation source profile of a coeval star cluster of mass $10^6$ $M_{\odot}$. We have chosen Kroupa initial mass function (\citealt{Kroupa2002}) where lower and upper cut-off mass are set to $0.1$ M$_{\odot}$ and $100$ M$_{\odot}$, Padova evolutionary tracks and metallicity Z$=0.4$ Z$_{\odot}$ (as in ST13). 

\begin{figure}
\centering
\includegraphics[height=3.1in,width=3.27in]{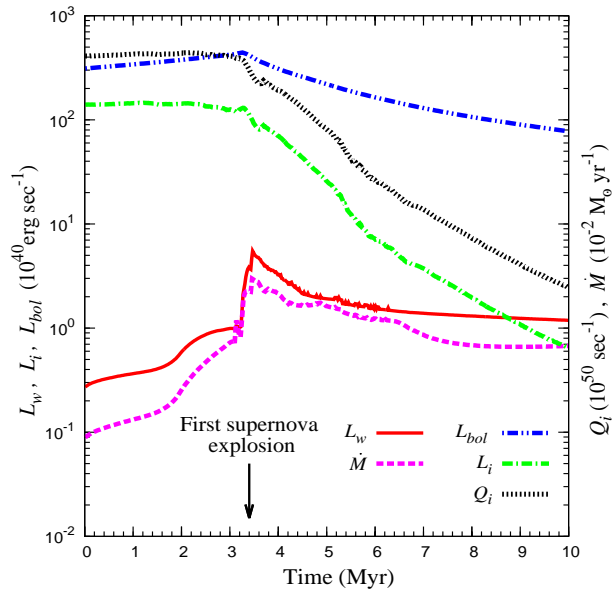} 
\caption{Star cluster output as a function of time. The solid red line and violet dashed line represent the mechanical luminosity ($L_{w}$) and mass-loss rate ($\dot{M}$). The blue dot-dashed, green dot-dashed and black dotted lines display the bolometric luminosity ($L_{bol}$), the ionizing (energy $> 13.6$ eV) photons luminosity ($L_{i}$) and flux ($Q_{i}$) respectively.}
\label{fig:outSB99}
\end{figure}

Figure \ref{fig:outSB99} displays the output of the Starburst99 synthesis model. The jump in the mechanical luminosity profile at $\approx 3.4$ Myr is due to the first supernova. After this, radiation luminosity ($L_{bol}$, $L_{i}$) and the ionizing photons flux ($Q_{i}$) fall down rapidly. It is worth noting that, after the first supernova, the mechanical luminosity becomes almost constant. This is because supernovae (SNe) become so frequent that on average, the winds power approaches a steady state.
\section{Analytical picture}
\label{sec:analytic}
Consider a scenario in which radiation from a central source (a star or star cluster) is incident on a neutral medium. The ionizing photons produce a fast moving ionization front (R-type). Ionization raises the gas temperature to $T\sim 10^4$ K and increases the particle number density, and hence the thermal pressure behind the ionization front is increased. As soon as the ionization front becomes D-type, the thermal pressure pushes the surrounding gas and piles it up into a thin shell (Ionization front-driven shocked ISM, shouldn't be confused with the wind driven shell); (for details see chapter $37$ in \citealt{Draine2011a}; \citealt{Haworth2015}; \citealt{Bisbas2015}; \citealt{Kim2016}). This is the case if we neglect the stellar and/or the supernova winds. In reality, the photons and winds interact with ambient medium simultaneously. Although the wind velocity is supersonic w.r.t the sound speed of ambient medium, its velocity is {\it less than} the velocity of R-type front, and so, the winds initially move in a medium which is already ionized. When the wind interacts with such a medium, it produces a shock and forms a wind-driven hot ($T\sim 10^7$ K) bubble. 

We consider a simple ISB model, where wind and radiation from the individual stars work cumulatively. We also consider a uniform-density interstellar medium (ISM) and therefore, we neglect all the effects which sustain density stratification. The structure of a wind-driven bubble with a constant mechanical luminosity is discussed in \citealt{Castor1975}, \citealt{Weaver1977} (for a detailed discussion see in \citealt{Bisnovatyi1995}). Assuming adiabatic evolution, the position of contact discontinuity (${R_{cd}}$), the thermal pressure in the SW region ($P_{sw}$) and the position of the reverse shock ($R_{rs}$) are
\begin{equation}
\label{eq:Rcd}
R_{cd} = \left(\frac{375(\gamma-1)}{28\pi(9\gamma-4)}\right)^{1/5}
         L_{w}^{1/5}\,\rho_{amb}^{-1/5}\,t_{dyn}^{3/5}
\end{equation}
\begin{equation}
\label{eq:Psw}
P_{sw} = 7\left(\frac{3(\gamma-1)}{700\pi(9\gamma-4)}\right)^{2/5}
         L_{w}^{2/5}\,\rho_{amb}^{3/5}\,t_{dyn}^{-4/5}
\end{equation}
\begin{equation}
\label{eq:Rrs}
R_{rs} = \left(\frac{L_{w}}{2\pi\, v_{w} P_{sw}}\right)^{1/2}
\end{equation}
where $L_{w}$ is the mechanical luminosity, $\rho_{amb}$ is the ambient density, $\gamma$ is the adiabatic index (chosen to be $5/3$) and $t_{dyn}$ is the dynamical time. These equations are valid as long as the system is adiabatic, i.e., when the dynamical time of the system is much shorter than the cooling time scale of various zones. In the adiabatic stage, the expansion of the bubble is determined by thermal energy of the SW region. This is known as Energy-dominated wind-driven bubble (ST13). Due to high density, the shell (shocked ISM) cools down and its cooling time scale (also see \citealt{MacLow1988}) is
\begin{equation}
\label{eq:tcoolshell}
\tau_{shell} = 2.7 \times 10^{4}\
              Z_{ISM}^{-25/59}\,n_{ISM}^{-42/59}\,L_{38}^{17/59}\ {\rm yr}
\end{equation}
where $L_{38}=L_{w}/10^{38}$ erg sec$^{-1}$, $n_{ISM}$ ($Z_{ISM}$) is the particle number density (metallicity) of the ambient medium. Shell cooling enhances the radiation pressure as the ionizing photons are trapped within it. A simple estimate of radiation force ($F_{rad}$) is
\begin{equation}
\label{eq:radmm}
F_{rad} = f_{trap} \frac{L_{bol}}{c}
\end{equation}
where $c$ is speed of light and $f_{trap}$ is the fraction of bolometric luminosity ($L_{bol}$) trapped within the shell. Numerical implementation of radiation force is discussed in section \ref{subsec:SemiRadHD}.

In addition to shell cooling, the interior of the bubble (shocked wind and free wind region) also lose energy, radiative cooling becomes important when the SW region is dominated by the mass evaporated due to thermal conduction. The radiative cooling time scale of the SW region (for details see section $2$ in \citealt{MacLow1988}) is
\begin{equation}
\tau_{sw} = 1.6 \times 10^{7}\
            Z_{ISM}^{-35/22}\,n_{ISM}^{-8/11}\,L_{38}^{3/11}\ {\rm yr}
\label{eq:tcoolsw}
\end{equation}
If the dynamical time becomes longer than this time scale, the SW region disappears and the free winds hit the shell directly and the expansion of bubble enters the momentum-dominated regime (\citealt{Martinez2014}; ST13). Using the above expressions, one can find the contribution of the radiation pressure in different regimes (see section $3$ in ST13).

The estimates in equations (\ref{eq:tcoolshell}) and (\ref{eq:tcoolsw}), however, depend on few crucial but probably invalid assumptions. First, the expressions for time scales (equation (\ref{eq:tcoolshell}) and (\ref{eq:tcoolsw})), have been calculated using the adiabatic bubble model. If $t_{dyn}\gtrsim\tau_{shell}$, then a significant amount of energy is lost within the shell and the expansion rate of the bubble becomes slower than in the adiabatic case. Second, although the expression for $\tau_{sw}$ does not show any direct dependence on the mass-loss rate $\dot{M}$, it assumes that the temperature of SW region ($T_{sw}$) lies between $10^{5}$ and $10^{7}$ K, and therefore, it indirectly contains the information about the wind velocity $v_{w}$ (because $T_{sw}\sim(m_{\rm H}/k_{B})\,v^2_{w}$), and hence the mass-loss rate ($\dot{M}\approx 2 L_{w}/v_{w}^2$) of the driving source. Therefore, the conclusions drawn using these cooling time scales may be off. The details about cooling time scales are discussed in sections \ref{subsec:adiabatic} and \ref{subsec:radvcool}.
\section{Simulation set-up}
\label{sec:simsetup}
In this section we describe the simulation set-up corresponding to results discussed in section \ref{sec:simresult}.
\subsection{Code settings}
\label{subsec:codeSetting}
We use publicly available code PLUTO (\citealt{Mignone2007}) to study the role of mechanical and radiation feedback on the ISM. We solve following set of hydrodynamic (HD) equations:
\begin{equation}
\label{eq:mass}
 \frac{\partial \rho}{\partial t}+\vec{\nabla}.(\rho\,\vec{v})  =  S_{\rho}        
\end{equation}
\begin{equation}
\label{eq:momentum}
 \frac{\partial }{\partial t} (\rho\,\vec{v} ) +\vec{\nabla}.(\rho\,\vec{v}\otimes\vec{v}) +\vec{\nabla}p  = \rho\,\vec{a}_{rad}      
\end{equation}
\begin{equation}
\label{eq:energy}
 \frac{\partial e}{\partial t}+\vec{\nabla}.\left[(e+p)\,\vec{v}\right] =  S_{e} - \vec{\nabla}.\vec{F}_{c}- q^{-} + q^{+} + \rho\,\vec{v}.\,\vec{a}_{rad}          
\end{equation}
Here $\rho$ is the mass density, $p$ the thermal pressure, $\vec{v}$ the fluid velocity, $e=\rho\,\epsilon+\,\rho\,v^{2}/2$ the total energy density and $\epsilon$ the specific thermal energy. The terms $S_{\rho}$ and $S_{e}$ in equation (\ref{eq:mass}) and (\ref{eq:energy}) are related to the mass-loss rate ($\dot{M}$) and the mechanical power ($L_{w}$) of the driving source, and therefore, they represent the mechanical source. Here, we are assuming that the wind energy is completely thermalized within the source region such that there is no additional mechanical momentum source term (\citealt{CC1985}). $\vec{F}_{c}$, $q^{-}$ and $q^{+}$ represent thermal conduction, cooling and heating respectively. The term $\vec{a}_{rad}$ refers to acceleration due to radiation pressure. 
\begin{table*}
 \centering
 \begin{tabular}{ l c c c c c c c c c} 
  \hline\hline 
 &  & \multicolumn{3}{c}{Source details} & \multicolumn{3}{c}{Physical processes} & \multicolumn{2}{c}{Simulation box details}\\
\cmidrule(l){3-5} \cmidrule(l){6-8} \cmidrule(l){9-10}                                     
Model &  $\rho_{amb}$   &   $L_{w}$ & $\dot{M}$ & R & H & C & T  & Box size (pc) & Grid points\\
        & \scriptsize{(${\rm m_{H}\,cm^{-3}}$)}    &  \scriptsize{(${\rm erg\,sec^{-1}}$)}    &  \scriptsize{(M$_{\odot}$ yr$^{-1}$)} & \scriptsize{(Rad. prs)}& \scriptsize{(Heating)} & \scriptsize{(Cooling)} & \scriptsize{(Thrm. cd.)}& [${\rm r_{min},r_{src},r_{max}}$] & [${\rm n_{src}, n_{rest}}$]   \\   
  \hline \vspace{.05in}
   ${\rm I0\_d3^{*}}$    & $10^{3}$ &   $10^{40}$  & $0.0315$  & -- & -- & --   & --  & --  & --\\
  
   ${\rm I1\_d3}$    & $10^{3}$ &   $10^{40}$  & $0.0315$  & -- & N & N   & N  & $[0.2,1.0,257.0]$  & $[16,2048]$\\
   ${\rm I1\_d3\_T}$    & $10^{3}$ &   $10^{40}$  & $0.0315$  & -- & N & N   & Y  & $[0.2,1.0,257.0]$  & $[16,2048]$\\
   \vspace{.05in}
   ${\rm I1\_d3\_CT}$   & $10^{3}$ &   $10^{40}$  & $0.0315$  & -- & N & Y   & Y  & $[0.2,1.0,161.0]$  & $[16,6400]$\\
  
   ${\rm I2\_d3}$    & $10^{3}$ &   $10^{40}$  & $0.0079$  & -- & N & N   & N  & $[0.2,1.0,257.0]$  & $[16,2048]$\\
   ${\rm I2\_d3\_T}$    & $10^{3}$ &   $10^{40}$  & $0.0079$  & -- & N & N   & Y  & $[0.2,1.0,257.0]$  & $[16,2048]$\\
   \vspace{.05in}
   ${\rm I2\_d3\_CT}$    & $10^{3}$ &   $10^{40}$  & $0.0079$  & -- & N & Y   & Y  & $[0.2,1.0,161.0]$  & $[16,6400]$\\
   
   ${\rm SB\_d3}$                & $10^{3}$ & sb99 & sb99 & N & N & N  & N   & $[0.2,1.0,257.0]$  & $[16,2048]$\\
   ${\rm SB\_d3\_T}$           & $10^{3}$ & sb99 & sb99  & N & N & N  & Y   & $[0.2,1.0,257.0]$  & $[16,2048]$\\
   ${\rm SB\_d3\_CT}$        & $10^{3}$ & sb99 & sb99 & N & N & Y  & Y   & $[0.2,1.0,161.0]$  & $[16,6400]$\\
   ${\rm SB\_d3\_HCT}$     & $10^{3}$ & sb99 & sb99 & N & sb99 & Y  & Y   & $[0.2,1.0,161.0]$  & $[16,6400]$\\
   \vspace{.05in}
   ${\rm SB\_d3\_RHCT}$  & $10^{3}$ & sb99 & sb99 & sb99 & sb99 & Y  & Y   & $[0.2,1.0,161.0]$  & $[16,6400]$\\
  \vspace{.05in}
   ${\rm SB\_d0\_RHCT}$  & $10^{0}$ & sb99 & sb99 & sb99 & sb99 & Y  & Y   & $[0.2,1.0,701.0]$  & $[16,17500]$\\
  \vspace{.05in}
   ${\rm SB\_d2\_RHCT}$  & $10^{2}$ & sb99 & sb99 & sb99 & sb99 & Y  & Y   & $[0.2,1.0,251.0]$  & $[16,6250]$\\
   ${\rm SB\_d4\_RHCT}$  & $10^{4}$ & sb99 & sb99 & sb99 & sb99 & Y  & Y   & $[0.2,1.0,101.0]$  & $[16,4000]$\\
 \bottomrule
  \end{tabular}
   \caption{Details of all models. In the extreme right column, ${\rm n_{src}}$ and ${\rm n_{rest}}$ denote the number of uniformly distributed grid points in the source region and in rest of the box. The symbol `T' stands for Thermal conduction, `C' for Radiative cooling, and `H' for Heating. The symbols `N' and `Y' indicate that the corresponding process is switched off and on respectively. The symbol `sb99' represents that it uses the output of Startburst99 (Figure \ref{fig:outSB99}), and also indicates that the corresponding set-up is on. `R' represents the radiation pressure. $\rho_{amb}$ is the initial density of the ambient medium. For all models, the initial temperature of the ambient medium is $T_{amb}=100$ K. Model labels I1, I2 stand for the idealized cases i.e. constant luminosity model. `SB' stands for the realistic model which uses the output of Starburst99. Therefore, label `${\rm SB\_d3\_HCT}$' represents a model which uses Starburst99 (${\rm SB\_}$), in which $\rho_{amb}=10^{3}$ ${\rm m_{H}\,cm^{-3}}$ (${\rm SB\_d3\_}$) and which includes thermal conduction, radiative cooling and heating.\\ $^{\rm *}$Label I0$\_$ stands for a bubble which follows analytical results as described in section \ref{sec:analytic}.}
 \label{table:model}
\end{table*}

Equations (\ref{eq:mass}) - (\ref{eq:energy}) are solved in spherical $1$-D geometry. Therefore, we can not study the effects of gas clumping and the opening of low density channels\footnote{The dominance of radiation pressure, as shown in section \ref{subsec:radiation}, may last shorter than obtained here because of the escape of the ionizing radiation through the low density channels.} in the swept-up ISM (these effects are discussed in \citealt{Harper2009}, \citealt{Dale2013}, \citealt{Rogers2013}). The simulation time step is limited by CFL number which is set to $0.4$ for numerical stability. Paying special attention to mass and energy conservation, we find that the best-suited solver for this problem is HLLC (\citealt{Toro1994}). In all our models, the simulation box starts from ${\rm r_{min}} = 0.2$ pc to ${\rm r_{max}}$, where the maximum box length ${\rm r_{max}}$ is chosen depending on the ambient density and physical processes. The details of all models are given in Table \ref{table:model}.

Note that all the terms on the right hand side of equations (\ref{eq:mass}) - (\ref{eq:energy}), are zero for ideal HD. Also note that, some of these terms have non-zero values depending on some physical criteria and the details are discussed in the following sections.

\subsection{Adding mechanical source}
\label{subsec:mechinput}
We consider the initial density ($\rho_{amb}$) and temperature ($T_{amb}$) of the ambient medium to be uniform and at rest. For later times, mass and energy are continuously added into a small region of radius $r_{src}$. Therefore, the mechanical source terms, $S_{\rho}=\dot{M}/V_{src}$ and $S_{e}=L_{w}/V_{src}$ (where $V_{src}=(4\pi/3) r^{3}_{src}$ is the volume of the source region) have non-zero value at $r\leq r_{src}$ (see equations (\ref{eq:mass}) and (\ref{eq:energy})). The radius of this region has been chosen such that the input energy rate is greater than the energy loss rate due to radiative cooling (\citealt{Sharma2014}, hereafter SRNS$14$), which gives
\begin{equation}
\label{eq:strongshock}
r_{src}\lesssim 5\ L^{1/3}_{40}\, \rho^{-2/3}_{amb,\,3}\, \Lambda^{-1/3}_{-24}\ {\rm pc}
\end{equation}
where $L_{40}=L_{w}/10^{40}$ erg sec$^{-1}$, $\rho_{amb,\,3}=\rho_{amb}/10^{3}$ $m{\rm_{H}}$ cm$^{-3}$ and $\Lambda_{-24}=\Lambda_{N}/10^{-24}$ erg cm$^{3}$ sec$^{-1}$ ($\Lambda_{N}$ is the normalized cooling function). In all our simulations, we set $r_{src}=$ $1$ pc and this is consistent with the radius for a star cluster of mass $\sim10^{6}$ M$_{\odot}$ (e.g., Figure 1 in \citealt{Murray2011}). 

Our mechanical source injection is similar to the model of \citealt{CC1985}, and therefore, we are assuming that the input energy is thermalized within the source region. It is worth noting that although adding momentum source term (kinetic energy model `KE' in SRNS$14$) is more realistic then a luminosity driven (LD) or thermal energy (TE) addition, both of them converge to similar profiles very quickly ($< 0.01$ Myr) for a compact (small ejecta radius ($r_{src}$)) and massive star cluster as we have taken; for details see section $5.2$ in SRNS$14$.
\subsection{Thermal conduction}
\label{subsec:conductionset}
The term ($-\vec{\nabla}.\vec{F}_{c}$) in equation (\ref{eq:energy}) represents thermal conduction, where $F_{c}$ is the conduction flux. As long as the electron mean free path ($\lambda_{m}$) is smaller than the temperature gradient length scale ($l_{T}$), the conduction flux can be defined as $\vec{F}_{c}= \vec{F}_{classical}= -\kappa \,\vec{\nabla} T$, where $\kappa$ is the coefficient of thermal conductivity (\citealt{Spitzer1962}). But if $\lambda_{m}>l_{T}$, then the definition of classical conduction breaks down and $|F_{c}|=F_{sat}\approx(2\,k_{B}\,T/\pi\,m_{e})^{1/2}\,n_{e}\,k_{B}\,T_{e}$  (\citealt{Cowie1977}). PLUTO deals with this by allowing a smooth transition between classical and saturated conduction fluxes\footnote{For details see the user's guide of \ufhref[webgreen]{http://plutocode.ph.unito.it/files/userguide.pdf}{PLUTO} and references therein.}. We evolve thermal conduction using Supper Time Stepping (STS) (\citealt{Alexiades1996}), and in classical regime, we use $\kappa=C\,T^{5/2}$ (\citealt{MacLow1988}).

Note that when the temperature is below $10^{4}$ K, atomic diffusion becomes important and $\kappa \approx C_{a}\,T^{1/2}$ (e.g., \citealt{Ferrara1993}). However, we find that thermal conduction for the gas of temperature $\lesssim 10^{4}$ K is negligible and the simulation results are independent of this choice. The effects of thermal conduction are not discussed explicitly in this paper; for details see SRNS$14$, \citealt{Weaver1977}.
%
\subsection{Cooling}
\label{subsec:coolset}
PLUTO uses operator splitting method to include the effect of radiative cooling (see the term $q^{-}$ in equation (\ref{eq:energy})). It solves
\begin{equation}
\label{eq:cooling}
 \frac{\partial (\rho\epsilon) }{\partial t} = -q^{-} 
\end{equation}
where $q^{-} = n_{i}\,n_{e}\,\Lambda_{N}(T, Z)$, $\Lambda_{N}$ is the normalised cooling function which is set to zero below 100 K (i.e., when the gas temperature $<T_{amb}$). We have confirmed that the final results do not depends on the assumption of $T_{amb}$ below $<1000$ K, corresponding to that of cold neutral medium. $n_{e}$ is electron number density and $n_{i}$ is ion number density. The numerical value of these quantities depends on metallicity and the ionization state of the gas. We use a special technique to account for the temperature dependence of the ionization state of the gas, for details, see Appendix \ref{app:cool-sim}. The tabulated cooling function is taken from PLUTO which has used CLOUDY (\citealt{Ferland1998}) to generate a normalized cooling table.
\subsection{Heating}
\label{subsec:heatset}
The dominant heating processes in our problem are photo-electric and photoionization heating, which are described as follows.
\subsubsection{Photo-electric heating}
We use the prescription of photo-electric (PE) heating rate per unit volume given by \citealt{Wolfire2003},
\begin{equation}
\label{eq:peheating}
n\,\Gamma_{PE}=1.3\times 10^{-24}\ \ n\,\epsilon\,G_{0} \ \ \ {\rm erg} \, {\rm cm}^{-3} \, {\rm sec} ^{-1}
\end{equation}
where $G_{0}$ is FUV radiation field normalised to Habing radiation, $n$ is the average number density of hydrogen nuclei and $\epsilon$ represents heating efficiency, which is approximated by a fit,
\begin{align}
\label{eq:peeffi}
\epsilon & = &\frac{4.9\times 10^{-2}}{1+4.0\times 10^{-3}(G_{0}T^{1/2}/n_{e}\phi_{PAH})^{0.73}}  \nonumber  \\
& & +  \frac{3.7\times 10^{-2}(T/10^{4})^{0.7}}{1+2.0\times 10^{-4}(G_{0}T^{1/2}/n_{e}\phi_{PAH})}
\end{align} 
Here $\phi_{PAH}$ is a parameter ($0.25\leqslant\phi_{PAH}\leqslant1.0$) which scales the electron-PAH collision rates and $n_{e}$ is the average number density of electrons. We have chosen $\phi_{PAH}=0.5$ as the standard value. 

\subsubsection{Photoionization heating}
The physics behind photoionization heating (PI) is not straightforward because it depends on the shape of the incident spectrum as well as on the ionization potential of individual elements. In present approach, we have considered photoionization only for hydrogen. Therefore, the photoionization heating rate per unit volume is
\begin{equation}
\label{eq:piheating}
n\,\Gamma_{PI} \approx \alpha_{B}\,n^2\,x^{2}\, E_{e}
\end{equation}
where $\alpha_{B}$ is `Case B' radiative recombination coefficient, $x\equiv n($H$^{+})/n=n_{e}/n$ is the ionization fraction of hydrogen atom and $E_{e}$ is the mean energy of photoelectrons. The above heating prescription assumes that hydrogen is nearly fully ionized (and ignores other elements) and that the number of the Lyman continuum photons absorbed in the ionized region are equal to the total number of recombinations to levels excluding the ground state (for details see section $27.1$ in \citealt{Draine2011a}). We have used temperature dependent $\alpha_{B}$ (see section $14.2$ in \citealt{Draine2011a}), and assume $E_{e}\simeq\left( \left<h\nu\right>_{i}-h\nu_{0}\right)$, where $\left<h\nu\right>_{i}\equiv L_{i}/Q_{i}$ is the average energy of the ionizing photons and $h\nu_{0}=13.6\,{\rm eV}$ is the threshold energy for photoionization from the ground state of hydrogen atom. Note that the numerical values of $L_{i}$ and $Q_{i}$ depend on the age of the cluster (Figure \ref{fig:outSB99}) and hence the term $\left<h\nu\right>_{i}$ is a function of time. 

To find $x$, we have used the condition for the photoionization balance i.e., the rate of `Case B' recombinations per unit volume is balanced by the rate of photoionization per unit volume. The photoionization balance condition gives 
\begin{equation}
\label{eq:pibalance}
\alpha_{B}\,n^2\,x^{2} = \frac{Q_{i}(r)}{4\pi\, r^{2}}\,n\,(1-x)\,\sigma_{pi}
\end{equation}
where $Q_{i}(r)$ is the rate at which the ionizing photons cross a spherical surface of radius $r$, $\sigma_{pi}=6.8\times 10^{-18}$ cm$^{2}$ is the photoionization cross-section of hydrogen. 

\begin{figure*}
\centering
\includegraphics[height=3.1in,width=6.5in]{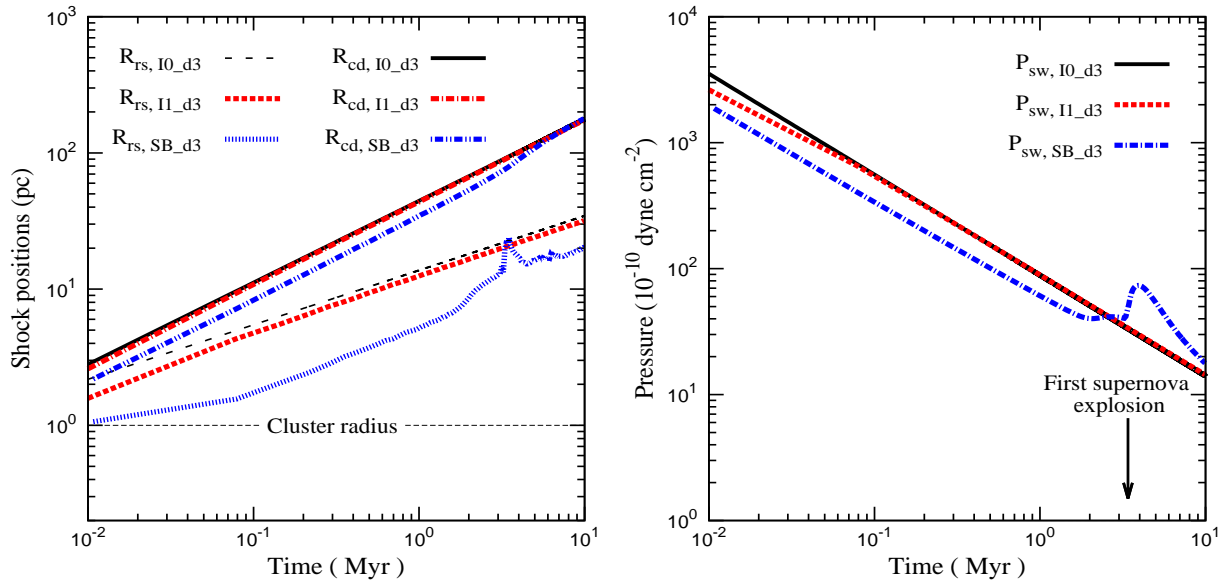} 
\caption{Comparison between constant luminosity and time-dependent luminosity model. Left panel displays the position of contact discontinuity ($R_{cd}$) and reverse shock ($R_{rs}$). Right panel shows the volume averaged thermal pressure ($P_{sw}$) in the SW region. In both panels, black colours represent analytical result (equations (\ref{eq:Rcd})-(\ref{eq:Rrs})) where $L_{w}=10^{40}$ erg sec$^{-1}$ and $\dot{M}=3.15\times10^{-2}$ M$_{\odot}$ yr$^{-1}$, red colours represent the idealised simulation using the same $L_{w}$ and $\dot{M}$, and blue colours show time-dependent luminosity model (Starburst99). For model details see Table \ref{table:model}.}
\label{fig:analycomp}
\end{figure*}

Note that the gas is heated up by the radiation field only when photons are able to interact with the gas. To consider this, we have introduced two attenuation factors $\phi_{n}$ and $\phi_{i}$ which represent the fraction of the FUV and EUV photons that able to reach at distance $r$ from the cluster. Therefore, we replace $G_{0}\rightarrow G_{0}\,\phi_{n}(r)$ and $Q_{i}(r)\rightarrow Q_{i}\,\phi_{i}(r)$. The details of $\phi_{n}$, $\phi_{i}$ are discussed in next section.

We add total heating rate per unit volume (i.e. sum of photoelectric and photoionization heating, $q^{+}=n\,\Gamma_{PE}\,+\,n\,\Gamma_{PI}$ for $r\gtrsim R_{cd}$) at the right hand side of equation (\ref{eq:cooling}), and therefore, for a given time, these two terms are calculated self-consistently.

Other than above mentioned heating processes, heating due to the X-ray photons (which are coming from the hot gas $\gtrsim10^{7}$ K) and cosmic-ray particles can also be important, but in the present approach, we have not considered them. This is because, in this problem, the X-ray luminosity ($L_{X}\sim 10^{36-37}$ erg sec$^{-1}$) is much lower than the luminosity of the input radiation source ($L_{bol}\sim10^{42}$ erg sec$^{-1}$, see Figure \ref{fig:outSB99}; also see section $2$ in \citealt{Pellegrini2011}) and moreover, the heating prescriptions of X-ray and cosmic-ray are strongly model dependent.
\subsection{Adding radiation source}
\label{subsec:SemiRadHD}

We use a phenomenological approach to include the effect of radiation pressure. From the discussion in section \ref{sec:analytic}, we note that the temperature of the free wind and SW regions are much higher than the ionization temperature of the gas and the gas acts as a plasma. The total column density of the free wind and SW region is $\ll 10^{25}$ cm$^{-2}$, and therefore, Thomson scattering is negligible. This makes the interior of the bubble almost transparent to the incident radiation. However, in the presence of dust, the shell need not be optically thin.

The radiation force per unit volume in the region $r\gtrsim R_{cd}$ can be written as
\begin{equation}
\label{eq:radsrc}
\rho\, a_{rad} = n\,\sigma_{d}\frac{L_{n} \phi_{n} + L_{i}\phi_{i}}{4\pi r^{2} c}
\end{equation}
Here $\sigma_{d}$ is dust opacity, $L_{n}$ and $L_{i}$ are the luminosities of non ionizing and ionizing photons respectively, $\phi_{n}$ and $\phi_{i}$ are the fraction of non-ionizing and ionizing photons that are able to reach a distance $r$. We have assumed $\phi_{i}\approx \phi_{n} = e^{-\tau}$ (for details see Appendix \ref{app:phi}), where $\tau=\int{n\,\sigma_{d}\,dr}$ is the dust absorption optical depth. We set $\sigma_{d}=10^{-21}$ cm$^{2}$ (\citealt{Draine2011b}) as the standard value, but we also consider different values of $\sigma_{d}$ to test the dependence of our results on it (see section \ref{subsec:diffsigma}). Note that, this approach is similar to \citealt{Draine2011b} except that, he has considered a static H{\scriptsize {II}} region, whereas by considering $a_{rad}$ as source term in equations (\ref{eq:momentum}) and (\ref{eq:energy}), we allow its evolution with time.
\section{Simulation results}
\label{sec:simresult}
In this section, we present the results from our simulations. The model parameters are summarised in Table \ref{table:model}. The verification of  the simulation set-up is confirmed by comparing\footnote{In addition, conservations laws have been confirmed for all models; for details see Appendix \ref{app:techdetails}.} with the analytical expressions and this is discussed in section \ref{subsec:adiabatic}. In section \ref{subsec:radvcool}, we discuss how the structure of the bubble differs from the adiabatic model in the presence of radiative cooling. Radiation can control the dynamics in two ways: through the heating and through radiation pressure. We have considered these two cases separately. The effect of heating is discussed in section \ref{subsec:heating}. We discuss the effects of radiation pressure on bubble dynamics in section \ref{subsec:radiation}.
\begin{figure*}
\centering
\includegraphics[height=3.1in,width=6.5in]{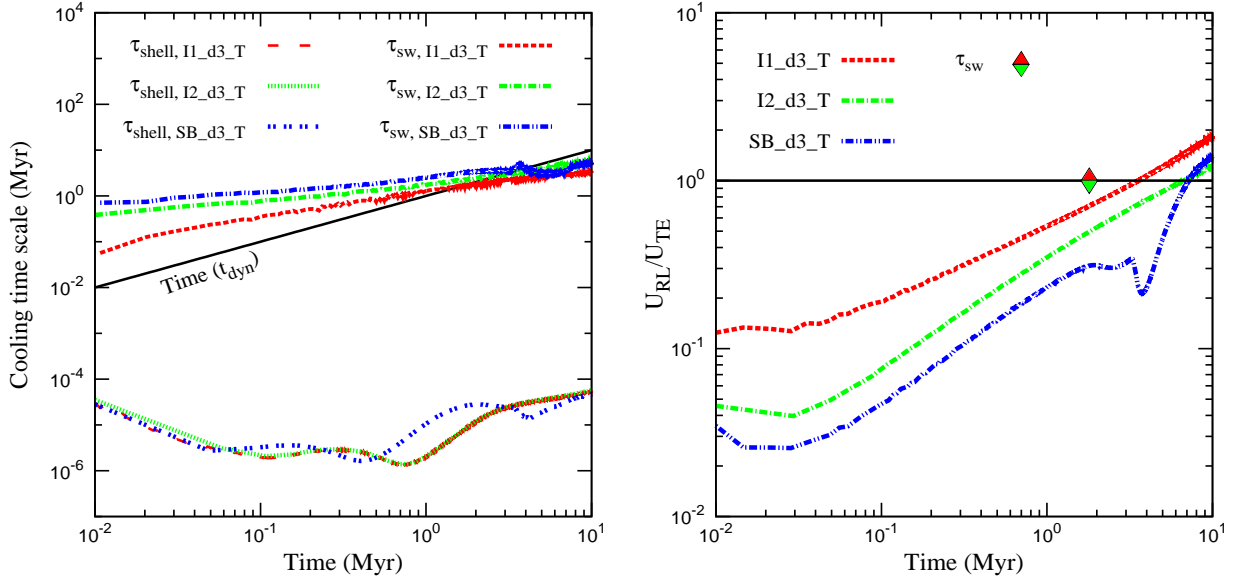} 
\caption{Cooling time scales of the shell and the SW region as a function of time. Black solid line in left panel represents the dynamical time ($t_{dyn}$) of the system. This panel shows that shell cooling time is much shorter than $t_{dyn}$, whereas cooling time scale of the SW region is close to $t_{dyn}$. Right panel displays the ratio of $U_{RL}$ to $U_{TE}$ (equations (\ref{eq:radloss}) \& (\ref{eq:therm})) for three different models as a function of $t_{dyn}$ which cross unity at $\approx 2.9$ Myr, $6.4$ Myr and $6.2$ Myr. The diamond mark represents $\tau_{sw}=1.8$ Myr (equation (\ref{eq:tcoolsw})).}
\label{fig:cooltime}
\end{figure*}
\subsection{Adiabatic model}
\label{subsec:adiabatic}
Here we study the difference between the constant luminosity and time-dependent luminosity model. We also discuss the cooling time scales of the shell and the SW region.
\subsubsection{Constant luminosity vs time-dependent luminosity}
\label{subsubsec:contVSvari}
A comparison between constant luminosity and time dependent luminosity runs is shown in Figure \ref{fig:analycomp}. Left panel displays the position of the contact discontinuity ($R_{cd}$) and reverse shock ($R_{rs}$). The volume averaged pressure in the SW is shown in the right panel. For both panels, black lines represent analytical results (equations (\ref{eq:Rcd}) - (\ref{eq:Rrs})) for constant luminosity. In the right panel, there is a hump in $P_{sw}$ at $\approx3.4$ Myr which is due to the first supernova (see Figure \ref{fig:outSB99}). This figure shows the differences between the constant luminosity and time-dependent luminosity. For both cases, we find that the analytical result matches with the simulation result even for the time-dependent luminosity (not shown in Figure \ref{fig:analycomp}).
\subsubsection{Cooling time scales}
\label{subsubsec:cooltimescales}
The cooling time scale is usually calculated by taking the ratio of the thermal energy to the instantaneous energy loss rate : $\tau_{cool}\approx k_{B}T/\left[(\gamma-1)\,n\,\Lambda_{N}\right]$, where $\Lambda_{N}$ is the normalised cooling function. Since, in this problem, the density and temperature of various zones vary with time, $\tau_{cool}$  also becomes a function of time. The time evolution of $\tau_{cool}$ for different models is shown in the left panel of Figure \ref{fig:cooltime}. It shows that the cooling time scale of the shell is much shorter than the dynamical time, and therefore, the shell can loss a significant amount of energy from an early time. However, for the SW region, $\tau_{sw}$ is close to the dynamical time, and because of this, it is difficult to conclude when the SW region becomes radiative.

\begin{figure*}
\centering
\includegraphics[height=7.29in,width=6.5in]{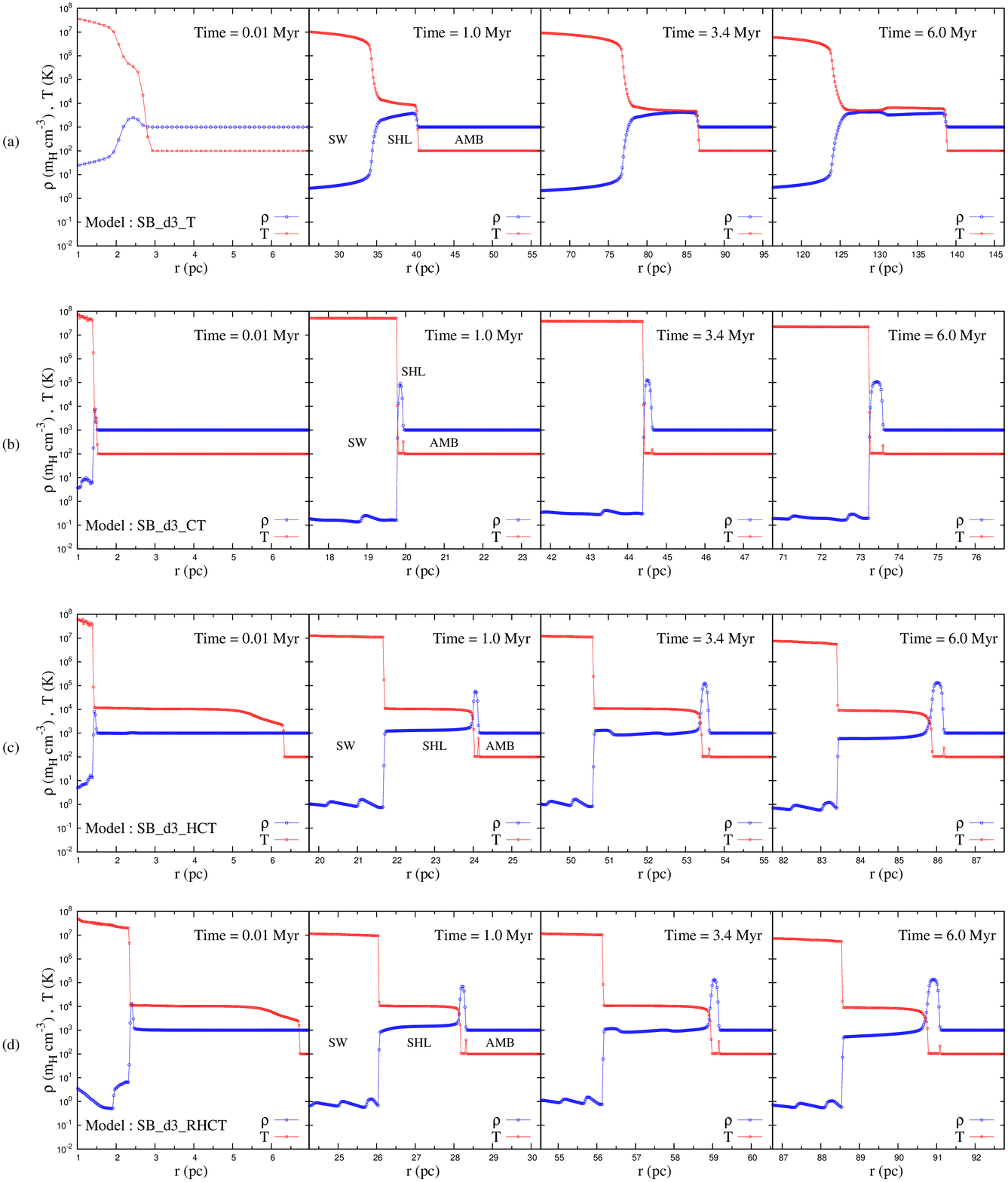} 
\caption{Snapshots of density and temperature profiles near the shell region in the presence/absence of different physical processes and radiation pressure at four different times : $t_{dyn}=$ $0.01$, $1.0$, $3.4$ and $6$ Myr. For all snapshots we show a zone extending $6$ pc, except for the last three snapshots of panel (a) which extends $30$ pc. Panel (a) shows adiabatic model (i.e., cooling is turned off, Model : ${\rm SB\_d3\_T}$) which is similar to Weaver bubble (\citealt{Weaver1977}), except that, here the mechanical source is a function of time. Panel (b) displays the shell structure in the presence of thermal conduction and radiative cooling (Model : ${\rm SB\_d3\_CT}$). Panel (c) shows the shell structure in the presence of thermal conduction, radiative cooling and heating (Model : ${\rm SB\_d3\_HCT}$). Panel (d) represents the shell structure for a realistic bubble in presence of radiation pressure, thermal conduction, heating and cooling (Model : ${\rm SB\_d3\_RHCT}$). The symbols SW, SHL and AMB denote the shocked wind region, shell and ambient medium respectively. A comparison of the shell structures of different panels at $0.01$ Myr shows that the radiation pressure launches the shock much faster than any other cases. At early times, the shell is transparent to the input radiation (e.g., see panels (c) and (d)), and a balance between heating and cooling keeps the temperature of the outside medium at $\sim 10^{4}$ K (also see Figure $4$ in \citealt{Martinez2014}).}
\label{fig:comprzoomin}
\end{figure*} 
\begin{figure*}
\centering
\includegraphics[height=3.1in,width=6.7in]{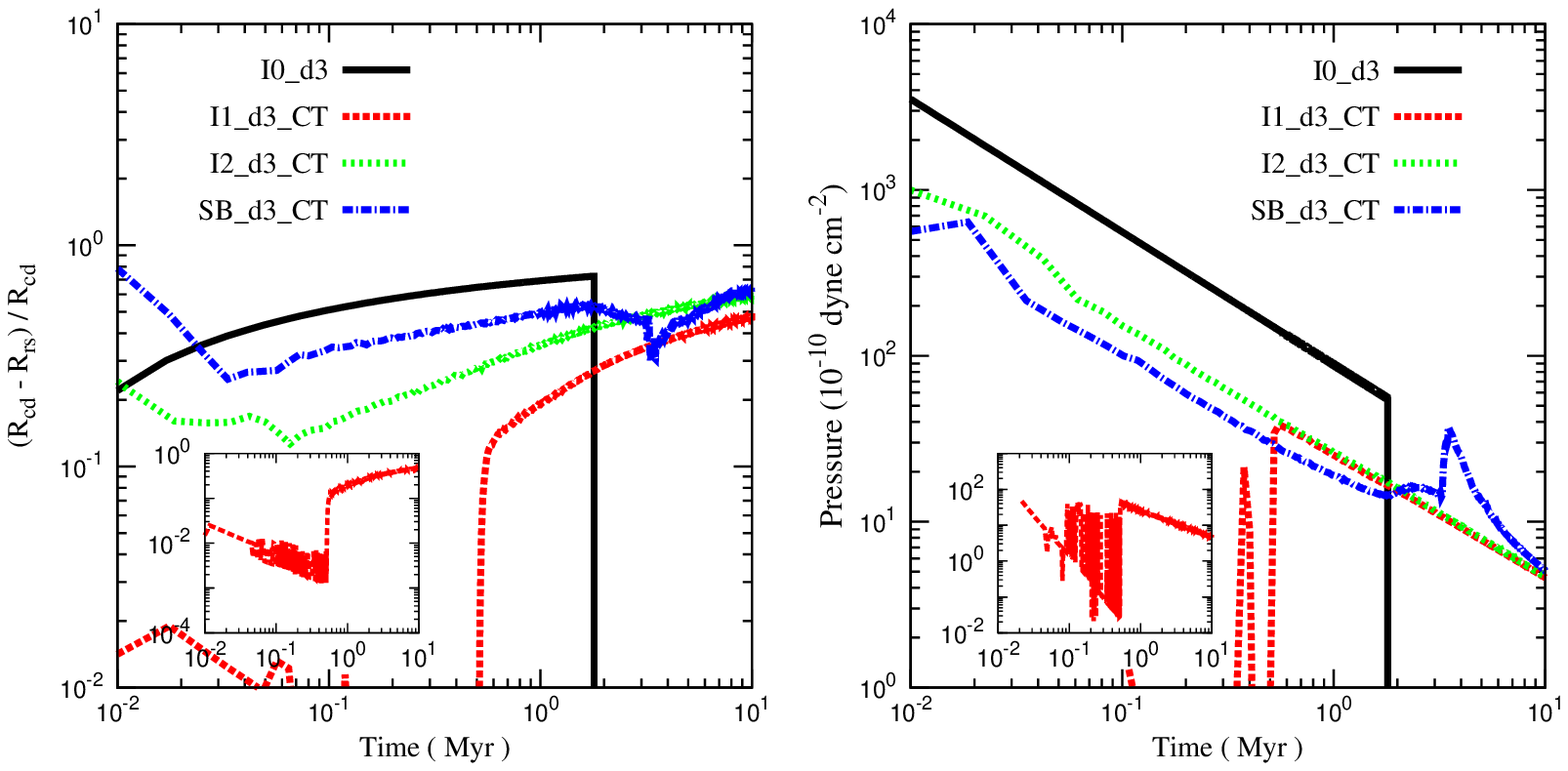} 
\caption{Time evolution of $\Upsilon_{sw}$ (equation (\ref{eq:upsilonSW}), left panel) and thermal pressure in the SW region ($P_{sw}$, right panel) for three different models (see Table \ref{table:model}). The black curves (in both panels) display the expected results from equation (\ref{eq:tcoolsw}) for the constant luminosity model, which shows the disappearance of the SW region at $t_{dyn}\gtrsim\tau_{sw}$ (see section \ref{sec:analytic}). Other curves represent the results from our simulations. The insets (in both panels) display the zoomed-out view of the red curve (i.e., large mass-loss rate), where horizontal and vertical axes represent the same variable as in respective panels. This figure shows that for high mass-loss rate (red curve, model I1\_d3\_CT), the excessive radiative losses cause a sawtooth behaviour in $(R_{cd}-R_{rs})/R_{cd}$ and in pressure ($P_{sw}$), and the reasons are discussed in section \ref{subsubsec:strucSW}.}
\label{fig:drprsSW}
\end{figure*}
One can also estimate the cooling time scale (as has been done to derive equation (\ref{eq:tcoolsw})) by estimating the total radiative losses until a given epoch. In this method, $\tau_{cool}$ of the SW region is obtained by equating the total energy loss ($U_{RL}$) from the SW region with the thermal energy ($U_{TE}$) in that region at that time. Therefore,
\begin{equation}
\label{eq:radloss}
 U_{RL} = \int^{t}_{t=0}\int^{R_{cd}}_{r=R_{rs}}dt\, d^{3}r\,\,\, n_{i}\,n_{e}\,\Lambda_{N}(T, Z)     
\end{equation}
\begin{equation}
\label{eq:therm}
 U_{TE} = \int^{R_{cd}}_{r=R_{rs}}d^{3}r\ \frac{p}{\gamma-1}     
\end{equation}
where $p$ is the thermal pressure. The radiative losses in the SW region are important when $U_{RL}/U_{TE}\gtrsim1$ (\citealt{MacLow1988}). The plot of $U_{RL}/U_{TE}$ as function of time for different models is shown in the right panel of Figure \ref{fig:cooltime}. The diamond mark represents the analytical result $\tau_{sw}\approx 1.8$ Myr which is obtained by using equation (\ref{eq:tcoolsw}). This figure shows that $U_{RL}/U_{TE}$ crosses unity at different times for models with different mass-loss rates but the same mechanical luminosity and ambient density, which means that the cooling time scale depends on the mass-loss rate. 

\subsection{Effects of radiative cooling}
\label{subsec:radvcool}
In this section, we study the effects of radiative energy loss on the shock structure.
\subsubsection{Structure of the Shell}
As expected from the discussion of the cooling time scale for the shell (see equation (\ref{eq:tcoolshell}) and Figure \ref{fig:cooltime}), the results of our simulations show that the shell is radiative right from the beginning. The snapshots of density and temperature profiles near the shell at different dynamical times are shown in panel (b) of Figure \ref{fig:comprzoomin}. This figure shows that the shell temperature is same as the ambient temperature (because, cooling function is set to zero below $100$ K, see section \ref{subsec:coolset}) and the width of the shell is small compared to the adiabatic case (see panel (a) in Figure \ref{fig:comprzoomin}). This indicates that the bubble is in isothermal phase. For an isothermal shock, the shell density is
\begin{equation} 
\label{eq:dshell}
\rho_{shell}\approx\mathcal{M}^2\,\rho_{amb}\ ,
\end{equation}
where $\mathcal{M}=\left(c_{s}/v\right)$ is the upstream isothermal Mach number, $c_{s}$ is the isothermal sound speed of the gas and $v$ is the velocity of the upstream materials (for details, see chapter $16$ in \citealt{Shu1992}). The shell width can be determined by equating the swept-up ambient mass with the shell mass which gives $\Delta R \approx \left(1/\mathcal{M}\right)^2 R_{cd}/3$.
\subsubsection{Structure of the SW region}
\label{subsubsec:strucSW}
To discuss the effects of radiative energy loss in the SW region, we define a parameter $\Upsilon_{sw}$
\begin{equation}
\label{eq:upsilonSW}
\Upsilon_{sw} = \frac{R_{cd}-R_{rs}}{R_{cd}}
\end{equation}
Therefore, $\Upsilon_{sw}\rightarrow0$ corresponds to the disappearance of the SW region. The plot of $\Upsilon_{sw}$ as a function of time is shown in Figure \ref{fig:drprsSW}. The solid black line in the left panel is obtained by using equations (\ref{eq:Rcd}) and (\ref{eq:Rrs}), and a sudden drop of $\Upsilon_{sw}$ is predicted by equation (\ref{eq:tcoolsw}). The other lines of this figure represent the results from simulations. The red curve refers to an arbitrary\footnote{If $N_{*}$ is the number of the stars that becomes supernova over the lifespan $t_{*}$, then the mechanical luminosity is $L_{w}\equiv (N_{*}/t_{*})E_{SN}$ and the mass-loss rate $\dot{M}=(N_{*}/t_{*})M_{SN}=(L_{w}/E_{SN})M_{SN}$, where $E_{SN}/M_{SN}$ is the average ejected energy/mass in each supernova. For a given $L_{w}=10^{40}\,\rm{erg\ sec^{-1}}$, $E_{SN}= 10^{51}\,\rm{erg}$, and $\dot{M}= 0.0315\, \rm{M_{\odot}\,yr^{-1}}$, $M_{SN}\approx 100\,\rm{M_{\odot}}$ which is extremely larger than the mass of OB stars.} large mass-loss rate, and is included here to illustrate the effects of excessive radiative cooling. This figure shows that the width of the SW region is {\it small} compared to the adiabatic model. This can be explained by the right panel which displays the pressure in the SW region in the presence of radiative cooling. In the presence of radiative cooling, a significant amount of thermal energy is radiated away, and as a result, thermal pressure in the SW region becomes less than in the adiabatic case. The position of reverse shock is determined by the balance between the thermal pressure of the SW  and the ram pressure of the wind. Thus, in the presence of radiative losses, the ram pressure pushes the reverse shock towards the shell, and the width of the SW region becomes smaller, but it does not disappear. We note that, the formation of the SW region depends on the mass-loss rate (e.g. red line) of  driving source which can be explained as follows.

\begin{figure}
\centering
\includegraphics[height=4.0in,width=3.0in]{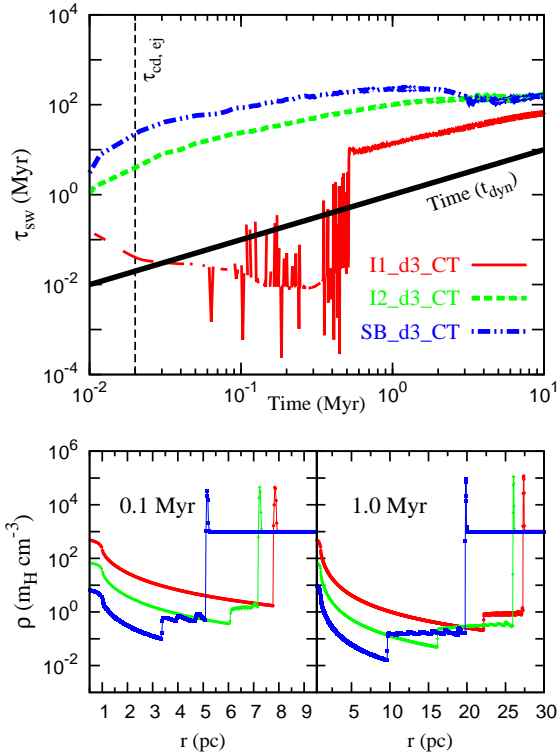} 
\caption{Upper panel displays the cooling time scale of the SW region (or pre-shocked wind layer) in the presence of radiative cooling as a function of time. The solid black line represents the dynamical time ($t_{dyn}$). The cooling time scale, which is shown in this panel, is similar to the left panel of Figure \ref{fig:cooltime}. However, it is worth noting that Figure \ref{fig:cooltime} shows cooling time scale in the {\it absence} of radiative cooling; this discrepancy arises because of the radiative energy loss from the dense shell. The vertical dashed line represents $\tau_{cd,ej}$ for red line (equation (\ref{eq:tcoolej})). Red line (in upper panel) shows spikes at early times because of the rapid cooling of the pre-shocked wind layer (also see Appendix \ref{app:resolution} which displays $\tau_{sw}$ as a function of time for different resolutions). Bottom panel displays the snapshot of density profiles at 0.1 Myr and 1.0 Myr. This panel shows that, for a high mass-loss rate (red line), the SW region is formed at late time. This figure also shows that the internal structure of ISB depends on the mass-loss rate of the driving source.}
\label{fig:coolzoom}
\end{figure}
Consider a scenario in which the ejecta material is accumulating near the contact discontinuity and the reverse shock just starts to form (i.e., $R_{rs}\approx R_{cd}$). At this moment, if a large fraction of the thermal energy is radiated away from the pre-shocked wind layer (or transient SW region which later appears as the SW region), then the thermal pressure becomes insufficient to overcome the force of the free wind (ram pressure) and hence, radiative cooling can suppress the formation of the SW region. Therefore, the SW region is formed if the radiative cooling time scale of the pre-shocked wind layer ($\tau_{ej,\,cd}$) is longer than the dynamical time. In presence of radiative cooling, the cooling time scale of the SW region (or pre-shocked wind layer) is shown in upper panel of Figure \ref{fig:coolzoom}. Bottom panel displays snapshot of density profiles at two different times. At early times, for the high mass-loss rate driving source (red line), the cooling time scale of the pre-shocked wind layer is shorter than the dynamical time, and therefore, it is strongly radiative.

To elaborate the above discussion, here we present an analytical criterion for the formation of the SW region by calculating the cooling time scale of the pre-shocked wind layer. We obtain an upper limit of the cooling time scale by assuming the density/pressure to be same as the free wind density/pressure at the contact discontinuity. Therefore, the cooling time scale is $\tau_{ej,\,cd} \approx P_{ej,\,cd}\,/\,[(\rho_{ej,\,cd}/m_{H})^{2}\,\Lambda_{N}/(\mu_{i}\mu_{e})]$, where $\rho_{ej,\,cd}\approx 0.056\ \dot{M}^{3/2}\,L^{-1/2}_{w}\,R_{cd}^{-2}$ and $P_{ej,\,cd}\approx 0.0106\ r^{4/3}_{src}\,\dot{M}^{1/2}\,L^{-1/2}_{w}\,R_{cd}^{-10/3}$ are density and thermal pressure of the free wind at the position of contact discontinuity (\citealt{CC1985}). This gives
\begin{equation}
\label{eq:tcoolej}
\tau_{ej,\,cd}\approx 2.31\ r_{src,\,1}^{20/9}\,L^{49/18}_{40}\ \rho^{-2/9}_{amb,3}\, \dot{M}^{-25/6}_{-2}\, \Lambda^{-5/3}_{-22}\,\, {\rm Myr}
\end{equation} 
where $r_{src,\,1}=r_{src}/1$pc, $\dot{M}_{-2}=10^{-2}$ M$_{\odot}$ yr$^{-1}$ and $\Lambda_{-22}=\Lambda_{N}/10^{-22}$ erg cm$^{3}$ sec$^{-1}$. 

In Figure \ref{fig:drprsSW} and \ref{fig:coolzoom}, red line corresponds to $\dot{M}_{-2}=3.15$ and $\tau_{ej,\,cd}\approx0.02$ Myr, and green line corresponds to $\dot{M}_{-2}=0.79$ and $\tau_{ej,\,cd}\approx6.2$ Myr. This indicates that if $\dot{M}$ is large, then the pre-shocked wind layer becomes radiative. This explains why at early time the SW region is absent for $\dot{M}_{-2}=3.15$ but it is present for $\dot{M}_{-2}=0.79$, although both have same $L_{w}=10^{40}$ erg sec$^{-1}$. At later time, $\rho_{fw}$ decreases (because, $\rho_{fw} \propto r^{-2}$), which increases the cooling time scale and when it becomes longer than $t_{dyn}$, then SW is formed. Once the SW region is formed, the temperature becomes so high ($\sim 10^{7}$ K) such that, $\Lambda_{N}$ drops to a small value $\lesssim 10^{-24}$ erg cm$^{3}$ sec$^{-1}$ which increases the cooling time scale almost abruptly (as shown in upper panel of Figure \ref{fig:coolzoom}). At later times, the SW region does not disappear. 

From the next section onward we drop the comparison with constant luminosity model and discuss the realistic bubble scenario which uses the output of Starburst99 (Figure \ref{fig:outSB99}).
\subsection{Effects of Heating}
\label{subsec:heating}
\begin{figure}
\centering
\includegraphics[height=2.7in,width=3.in]{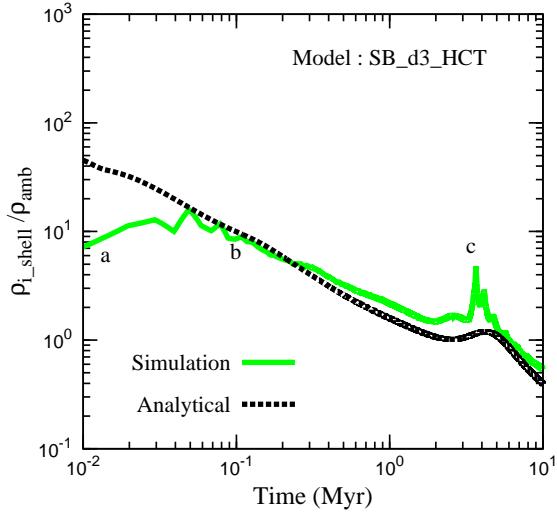} 
\caption{Density of the ionized region of the shell as a function of time. Black dotted line shows the analytical result using equation (\ref{eq:dishell}), where $\mathcal{M}$ is calculated from the simulation, and the green solid line shows the volume averaged density in the ionized part of the shell which is obtained from the simulation. Labels a, b and c are discussed in section \ref{subsec:heating}.}
\label{fig:denshelio}
\end{figure}
For a realistic evolution, we should consider heating in addition to cooling (for set-up details, see section \ref{subsec:heatset}).

The results of heating (plus cooling) are shown in panel (c) of Figure \ref{fig:comprzoomin}. By comparing panels (b) (without heating) and (c) (with heating), we notice that heating diminishes the effect of radiative energy loss (see the horizontal axis which denotes the radial coordinate) and also changes the structure of the shell. The part of shell facing the cluster has a temperature $\approx 10^{4}$ K ($T_{i}$), whereas the temperature of outer part is  $10^{2}$ K ($\approx T_{amb}$). This is because, when radiation passes through a dusty medium, the dust absorbs radiation flux and  does not allow it to propagate further. The neutral part of the shell is kept isothermal with the ambient medium and the density jump can be found by using equation (\ref{eq:dshell}). The density of ionized part of the shell can be found by assuming pressure balance between ionized ($P_{i\_shell}$) and neutral part ($P_{n\_shell}$) of the shell and this gives
\begin{equation} 
\label{eq:dishell}
\frac{\rho_{i\_shell}}{\rho_{amb}}\approx\mathcal{M}^2\left(\frac{T_{amb}\,\mu(T_{i})}{T_{i}\,\mu(T_{amb})}\right)=1.21\ v^2_{s,1}\,\, T^{-1}_{i,4}\,\, \mu(T_{i})\,
\end{equation}
where $\mu(T)$ is the mean mass per particle in the gas at temperature $T$ (see Appendix \ref{app:cool-sim}), $v_{s, 1}=(v_{s}/10\,{\rm km\,sec^{-1}})$ is the shock velocity and $T_{i,4}=(T/10^{4}\,{\rm K})$. Figure \ref{fig:denshelio} shows the ratio of the volume averaged density in the ionized region of the shell ($\rho_{i\_shell}$) to the ambient density ($\rho_{amb}$). This figure shows that equation (\ref{eq:dishell}) holds well only at late time. This can be explained as follows.

The density jump at time $t_{a}$ (time corresponding to label `a') is $\approx 8$, which indicates that the shell is radiative from the early time, but the neutral layer of the shell is not formed till $t_{b}$ (see panel (c) in Figure \ref{fig:comprzoomin}). For the time between $t_{b}$ and $t_{c}$, the simulation result follows equation (\ref{eq:dishell}) approximately, but equation (\ref{eq:dishell}) slightly underestimates its value because $P_{i\_shell}>P_{n\_shell}$. At $t_{c}$, there is a jump in $(\rho_{i\_shell}/\rho_{amb})$ because the mechanical energy suddenly increases at that time (Figure \ref{fig:outSB99}) which pushes the contact discontinuity.
\begin{figure}
\centering
\includegraphics[height=3.01in,width=3.29in]{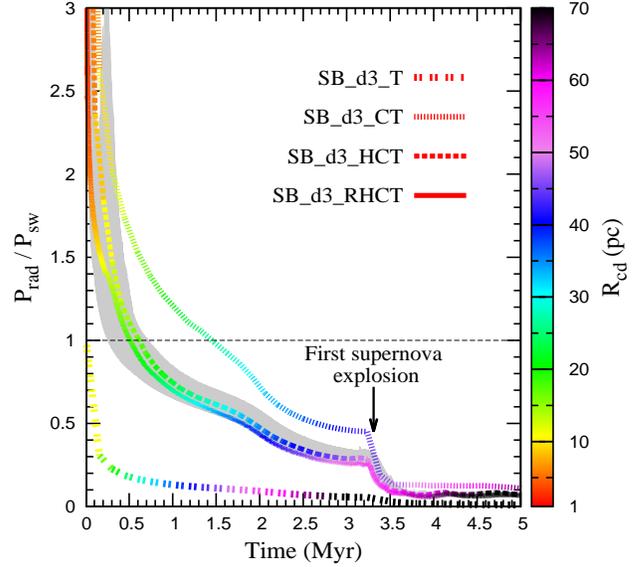} 
\caption{The ratio $\epsilon_{rad}$ ($\equiv P_{rad}/P_{sw}$) of radiation pressure ($P_{rad}$) to thermal pressure of the SW region ($P_{sw}$) as a function of time in the presence/absence of different physical processes. The coloured palette at rhs displays the position of the shell ($R_{cd}$) at a given epoch. The shaded region shows the range of $\epsilon_{rad}$ for $1\lesssim\rho_{amb}\lesssim10^{4}$ $m\rm{_{H}\,cm^{-3}}$ for the runs SB\_d0\_RHCT, SB\_d2\_RHCT, SB\_d3\_RHCT and SB\_d4\_RHCT (Table \ref{table:model}). The horizontal dashed line corresponds to $\epsilon_{rad}=1$. This figure shows that for a realistic bubble, $\epsilon_{rad}$ is weakly sensitive to the  ambient density.}
\label{fig:ratioradprs}
\end{figure}
\subsection{Effects of radiation pressure}
\label{subsec:radiation}
Radiation pressure on the shell is defined as 
\begin{equation}
\label{eq:radprs}
P_{rad} = f_{trap}\frac{L_{bol}/c}{4 \pi R^{2}_{cd}}
\end{equation}
where $f_{trap}$ is the trapping fraction of bolometric luminosity which is chosen to be unity. Using equations (\ref{eq:Rcd}) and (\ref{eq:Psw}), the
ratio of the radiation pressure ($P_{rad}$) to thermal pressure ($P_{sw}$) can be written as
\begin{equation}
\label{eq:ratioPrPsw}
\epsilon_{rad}=\frac{P_{rad}}{P_{sw}}\simeq 0.016\ \rho^{-1/5}_{amb,3}\,\, \eta^{-4/5}_{mech}\,\, L^{-4/5}_{w,40}\,\, L_{bol,42}\,\, t^{-2/5}_{6}
\end{equation}
Here we have replaced $L_{w}$ by $(\eta_{mech}\,L_{w})$, where $\eta_{mech}\,$ is the mechanical efficiency of the superbubble and $t_{6}=\left(t_{dyn}/10^6\,{\rm yr}\right)$. For adiabatic case $\eta_{mech}=1$, however for realistic bubble, $\eta_{mech}$ ($<1$) depends on $\rho_{amb}$, and also on heating and cooling (a general definition of $\eta_{mech}$ is given in section \ref{subsec:cooldiag}). The time evolution of $\epsilon_{rad}$ is shown in Figure \ref{fig:ratioradprs}. The coloured palette on right side of this figure shows the position of contact discontinuity at that epoch. From adiabatic bubble model, $\epsilon_{rad}$ is expected to follow the bottom dashed line in Figure \ref{fig:ratioradprs}. The shaded region shows $\epsilon_{rad}$ for $1\lesssim\rho_{amb}\lesssim10^{4}$ $m\rm{_{H}\,cm^{-3}}$. Therefore, the result of {\it realistic} simulation shows that the radiation pressure dominates over thermal pressure of the SW region {\it before} $\lesssim1$ Myr.

Note that, $\epsilon_{rad}$ is almost insensitive to the ambient density (Figure \ref{fig:ratioradprs}). Therefore, at a given epoch, $\epsilon_{rad}$ is roughly proportional to $ L^{1/5}_{bol}$ (see equation (\ref{eq:ratioPrPsw})), because the ratio of $L_{bol}$ to $L_{w}$ does not depend on the mass of the star cluster (\citealt{Leitherer99}). Hence, the role of radiation pressure is important for the massive star clusters (see Appendix \ref{app:Mcluster}).
\begin{figure}
\centering
\includegraphics[height=3.in,width=3.in]{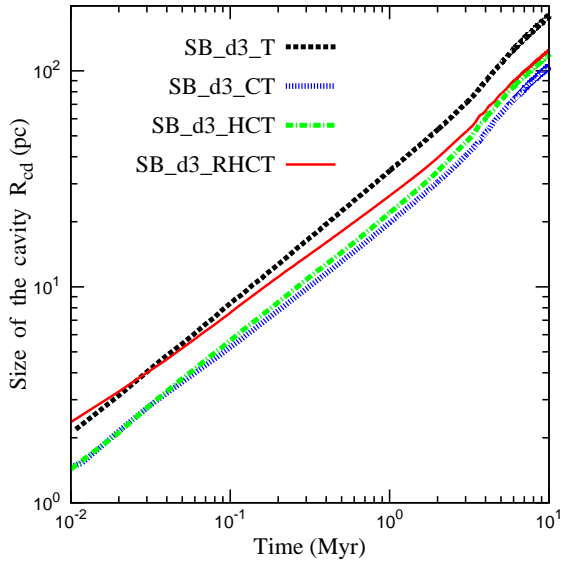} 
\caption{The size of the bubble in presence/absence of different physical processes. This figure shows that the radiation pressure is important only at early times.}
\label{fig:radeff}
\end{figure}

In the presence of radiation pressure (for set-up details see section \ref{subsec:SemiRadHD}), the snapshots of density and temperature profile at different times are shown in panel (d) of Figure \ref{fig:comprzoomin}. By comparing the shell structure at $0.01$ Myr in different panels, we find that the radiation pressure helps to launch the shock into the ISM at early times. Figure \ref{fig:radeff} shows the size of the cavity ($R_{cd}$) in presence/absence of different physical processes and radiation pressure. The effect of radiation on the dynamics of ISB is important at an early time, but as time evolves, ISB slowly makes transition from the radiation pressure dominated regime and enters into the thermal pressure dominated regime.

\section{Discussion}
\label{sec:modlobscomp}
Most of the results discussed in previous sections are based on a fixed ambient density model ($\rho_{amb}=10^{3}$ $m\rm{_{H}\,cm^{-3}}$) and single opacity parameter $\sigma_{d}=10^{-21}$ cm$^{2}$ (\citealt{Draine2011b}). In sections \ref{subsec:diffsigma}, \ref{subsec:ddiagram}, we first explore the dependence of simulation results on those parameters, and then we discuss the energetics of the superbubbles in section \ref{subsec:cooldiag}. We compare our results with other models, and with observations of 30 Doradus in sections \ref{subsec:unem} and \ref{subsec:othermodlcomp}.
\subsection{The choice of dust absorption coefficient}
\label{subsec:diffsigma}
\begin{figure}
\centering
\includegraphics[height=3.in,width=2.9in]{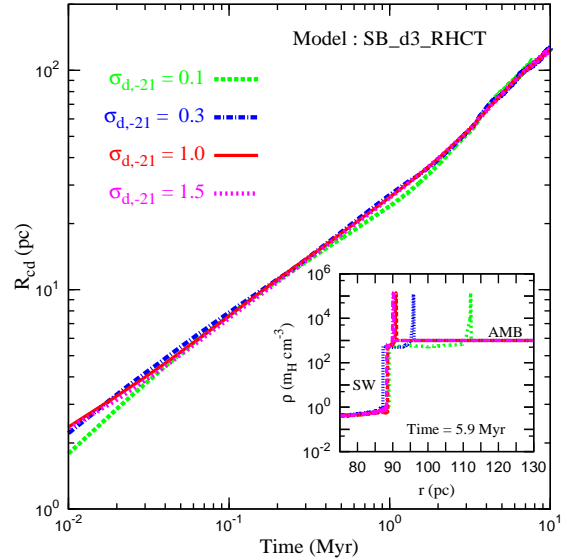} 
\caption{The dependence of bubble size on the dust absorption coefficient ($\sigma_{d}$). This figure displays the position of contact discontinuity ($R_{cd}$) as a function of time. Inset shows the density profile near the shell region for different values of $\sigma_{d, -21}$ where $\sigma_{d,-21}=(\sigma_{d}/10^{-21})$ cm$^{2}$.}
\label{fig:diffk}
\end{figure}

The dust absorption coefficient ($\sigma_{d}$) within ISM is not well characterised because it depends on various factors such as grain size distribution, dust-to-gas ratio, wavelength of the incident radiation etc. The dependence of our results on the choice of $\sigma_{d}$ is shown in Figure \ref{fig:diffk} which indicates that the position of contact discontinuity is almost independent on the choice of $\sigma_{d}$. The inset shows that the width of the photoionized region increases with a decrease in $\sigma_{d}$. This can be explained as follows. 
\begin{figure}
\centering
\includegraphics[height=5.6in,width=2.9in]{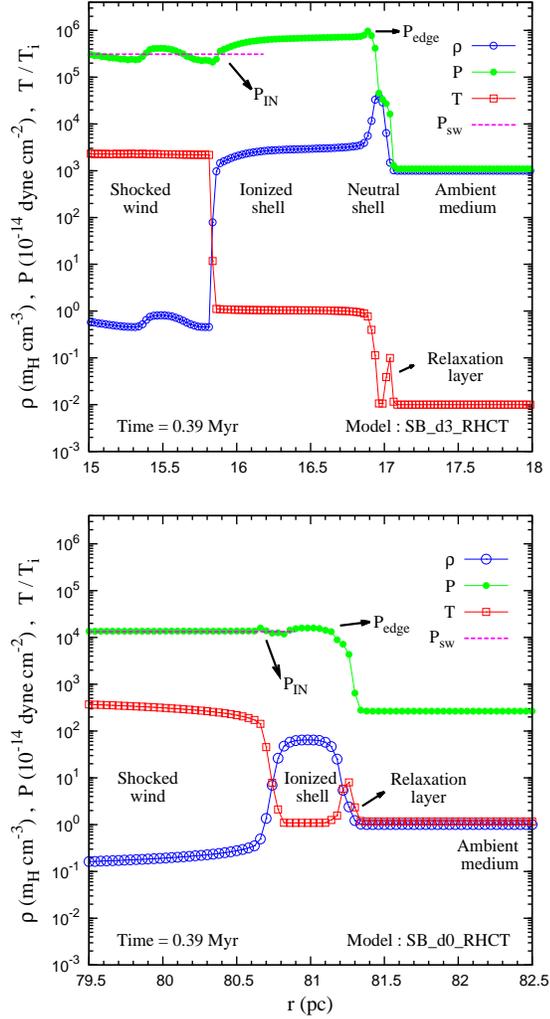} 
\caption{Snapshots of density ($\rho$), pressure ($P$) and the normalized temperature (w.r.t $T_{i}=10^{4}$ K) near the shell at $0.39$ Myr for the runs SB\_d3\_RHCT (top panel) and SB\_d0\_RHCT (bottom panel). $P_{sw}$ is the volume averaged pressure in the SW region at that time. The top panel shows that for a high density ambient medium (SB\_d3\_RHCT), the shell is partially ionized. Bottom panel shows that for a low density ambient medium (SB\_d0\_RHCT), the shell is completely ionized. For the model SB\_d3\_RHCT : $P_{sw} \approx 30.97\times 10^{-10}\, \rm{cgs},\, R_{cd} \approx 15.83\,\rm{pc},\, P_{edge} \approx 96.77\times 10^{-10}\, \rm{cgs} ,\, R_{edge} \approx 16.88\, \rm{pc}$, which gives $\Omega \simeq 0.36$. Similarly, for SB\_d0\_RHCT : $P_{sw} \approx 1.32\times 10^{-10}\, \rm{cgs},\, R_{cd} \approx 80.65 \, \rm{pc},\, P_{edge} \approx 1.43\times 10^{-10}\, \rm{cgs},\, R_{edge} \approx 81.14 \, \rm{pc}$, which gives $\Omega \simeq 9.7$. These values of $\Omega$ are shown in Figure \ref{fig:yeh}.}
\label{fig:prsdentemp}
\end{figure}

A larger $\sigma_{d}$ corresponds to a stronger radiation pressure ($P_{rad}$) at the inner layer of the shell (i.e., $f_{trap}\rightarrow 1$) but it decreases the heating efficiency because of the optical depth ($\tau$) which diminishes its strength by a factor $e^{-\tau}$. A lower value of $\sigma_{d}$ decreases the strength of $P_{rad}$ at the inner layer, but it increases heating efficiency which reflects on the shell structure. 
\subsection{Different regimes in a diagnostic diagram}
\label{subsec:ddiagram}

We use the dimensionless diagnostic diagram suggested by \citealt{Yeh2012} (hereafter YM12) to identify the dominant feedback mechanism (see Figure 1 in YM12). YM12 suggested two dimensionless parameters $\Omega$ and $\Psi$. The parameter $\Psi=R_{IF}/R_{ch}$, where $R_{IF}$ is the location of the ionized shell and $R_{ch}$ is the characteristic radius of the standard Str\"omgren sphere ($R_{st}$) at where the gas pressure is equal to the total unabsorbed radiation pressure $L_{bol}/(4\pi R^{2}_{st} c)$. The expression of $R_{ch}$ is given as,
\begin{equation}
R_{ch} = \frac{\alpha_{B} L^{2}_{bol}}{12 \pi \left(k_{B}T_{i}c\mu_{i}/\mu\right)^2 Q_{i}} \,
\end{equation}
where $L_{bol}$ is the bolometric luminosity and $Q_{i}$ is flux of the ionizing photons, $\mu$ and $\mu_{i}$ are the mean mass per atom and the mean mass per ion respectively, $T_{i}$ is temperature of ionized medium. Since $L_{bol}$ and $Q_{i}$ depend on the age of a cluster, the numerical value of $R_{ch}\sim 30\ L^{2}_{bol,42}\,Q^{-1}_{i,52}\,T^{-2}_{i,4}$ pc is a function of time.
According to this definition, if $\Psi>1$ (i.e., $R_{ch}<R_{IF}$), the bubble expands either in standard H{\scriptsize {II}} regime (Str\"omgren sphere) or wind dominated regime, and if $\Psi<1$ then the size of the bubble is smaller than the standard case. Therefore, $\Psi$ is a measure of compactness of H{\scriptsize {II}} region.
\begin{figure}
\centering
\includegraphics[height=3.in,width=3.4in]{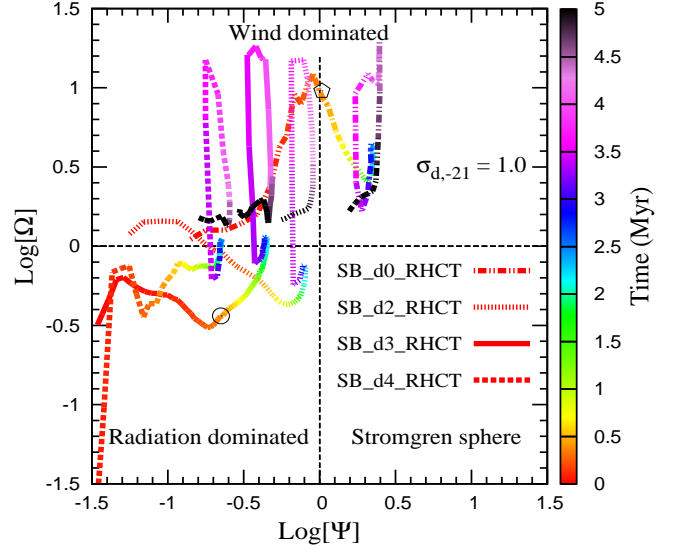} 
\caption{Time evolution of the diagnostic parameters in $\Omega-\Psi$ parameter space for four different ambient densities but for a fixed dust opacity $\sigma_{d}=10^{-21}$ cm$^{2}$. The coloured palette on right-hand side of this figure represents the dynamical time. The marked circle and pentagon denote $\Omega$ for the models SB\_d3\_RHCT and SB\_d0\_RHCT  at $0.39$ Myr respectively (see Figure \ref{fig:prsdentemp}).}
\label{fig:yeh}
\end{figure}

The second parameter $\Omega$ is defined as
\begin{equation}
\label{eq:omega}
\Omega=\frac{P_{IN}V_{IN}}{P_{IF}V_{IF}-P_{IN}V_{IN}}
\end{equation}
where $P_{IN}$ is the pressure at the inner edge (i.e., the edge of shell facing the driving source) of the ionized shell and $P_{IF}$ is the pressure at the ionization front (IF). Therefore, $P_{IF}V_{IF}-P_{IN}V_{IN}$ represents the difference in the product of the pressure and volume between ionization front and inner edge of the ionized shell. In our case, $R_{IN}\approx R_{cd}$ and $P_{IN}\approx P_{sw}$, and $P_{IF}$ is the pressure at outer edge of the ionized shell (i.e., $R_{edge}\approx R_{IF}$ and $P_{IF}\approx P_{edge}$), therefore $\Omega=P_{sw}R^3_{cd}/\left[P_{edge}R^3_{edge}-P_{sw}R^3_{cd}\right]$  (\citealt{Martinez2014}). To illustrate the significance of these parameters see Figure \ref{fig:prsdentemp}. In thin shell limit, $\Omega\approx 1/(P_{edge}/P_{sw}-1)$, where $P_{edge}$ can be simply assumed to be $P_{edge}\approx P_{sw}+P_{rad}$ and therefore, Log[$\Omega$]$<0$ represents radiation dominated regime and Log[$\Omega$]$>0$ represents wind/thermal pressure dominated regime. It is worth mentioning that, for the realistic case, $R_{edge}\neq R_{cd}$, and we find that $P_{edge}$ depends not only on the radiation pressure but also on heating and column density of the shell. Therefore, although we have seen that $\epsilon_{rad}=P_{rad}/P_{sw}$ depends weakly on the ambient density (Figure \ref{fig:ratioradprs}), here we find that $\Omega$ is density dependent (Figure \ref{fig:prsdentemp}). 

The evolutionary tracks of $\Omega$ for four different $\rho_{amb}$ but for a fixed $\sigma_{d}$ are shown in Figure \ref{fig:yeh}. The coloured palette on right-hand side of this figure represents the dynamical time. We see a similar evolution for different $\sigma_{d}$ (not shown in Figure \ref{fig:yeh}). At early times, the size of the bubble is much smaller than $R_{ch}$ ($\approx 70$ pc) i.e., $\Psi<1$. With time the bubble size increases, and therefore, $\Psi$ keeps increasing until the first supernova. After that, $Q_{i}$ falls so rapidly that $R_{ch}$ ($\gtrsim 175$ pc) increases faster than the bubble, and $\Psi$ starts to decrease. Note that, for high density media ($\rho_{amb} \gtrsim 10^{2}$ $m{\rm_{H}\,cm^{-3}}$), $\Psi$ is always less than unity. This figure shows that, for high density media, the bubble moves into the radiation dominated regime (i.e., Log[$\Omega$]$<0$) from an early time ($\lesssim 0.1$ Myr). It makes transition to the thermal/wind dominated regime after $\gtrsim 3$ Myr (i.e., Log[$\Omega$]$\gtrsim 0$) which corresponds to the epoch of steep decrease in $P_{rad}/P_{sw}$ in Figure \ref{fig:ratioradprs}. For a low density medium ($\rho_{amb}\approx1$ $m{\rm_{H}\,cm^{-3}}$), and bubbles always remain in the thermal pressure/wind dominated regime.

Note that, the diagnostic parameter $\Omega$ slightly overestimates the radiation dominated regime because the shell is not geometrically thin ($R_{edge}\neq R_{cd}$). In reality, the radiation pressure dominating regime ends at $\lesssim1$ Myr (see Figure \ref{fig:ratioradprs}).

\subsection{Temperature distribution of cooling losses \& the retained energy}
\label{subsec:cooldiag}
\begin{figure}
\centering
\includegraphics[height=1.9in,width=3.2in]{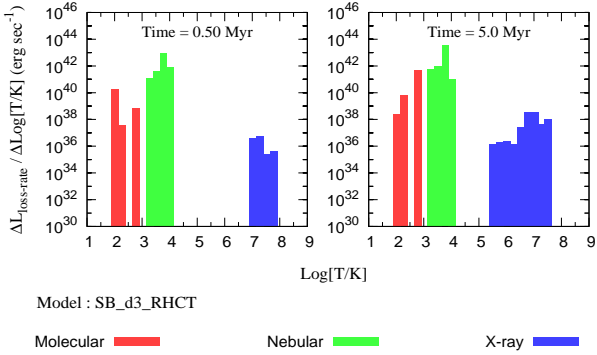} 
\caption{Temperature distribution of cooling losses at 0.5 Myr and at 5.0 Myr. Cooling losses are confined to three different temperature bands which are denoted by red colour for molecular emission ($\lesssim 10^{3}$ K), green colour for nebular emission ($10^3\lesssim T\lesssim 10^5$ K) and blue colour for X-rays ($T\gg 10^{5}$ K). This figure shows that most of the radiative losses occur at $\sim10^{4}$ K, with sub-dominant losses at $\lesssim10^3$ K and $\sim10^{6}-10^{8}$ K.}
\label{fig:cooltempdis}
\end{figure}
To compare with observations, we calculate the radiative output of our superbubble (model SB\_d3\_RHCT) at various temperatures. We create logarithmic bins in temperature ($\Delta {\rm Log} [T/K]$) and calculate the total radiative losses per unit time in each temperature bin. Figure \ref{fig:cooltempdis} shows that radiative losses occur at molecular ($\lesssim 10^{3}$ K), nebular ($\sim 10^4$) and X-ray temperatures ($\sim 10^6-10^8$ K). The molecular radiation comes from the radiative relaxation layer ahead of the dense shell (see the top panel of Figure \ref{fig:prsdentemp}). The nebular emission comes from the $\sim 10^4$ K shell and X-rays come from the shock heated, conductively mass-loaded shocked wind. A similar temperature distribution of luminosity is seen for most of our models. The highest luminosity comes from the gas at $\sim 10^4$ K, which should emit in nebular lines and continuum. The luminosity in the nebular temperature band is $\sim 10^{42}$ erg sec$^{-1}$, comparable to the ionizing luminosity from driving source. A significant fraction of optical and X-ray emission is expected to be absorbed by the large column density material in the shell. In reality, the radiation leaks out because of the clumpiness in the shell. The X-ray luminosity is $\sim 10^{36-37}$ erg sec$^{-1}$, comparable to the observed X-ray luminosity of $30$ Doradus (\citealt{Wang1991,Townsley2006}).
\begin{figure}
\centering
\includegraphics[height=3.2in,width=2.99in]{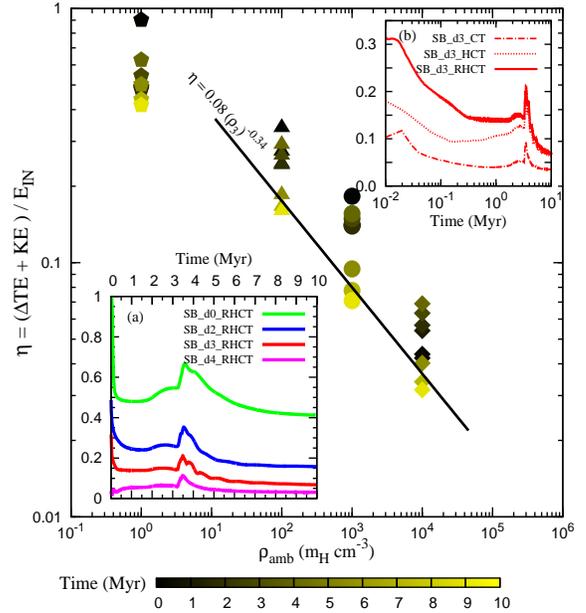} 
\caption{Fraction of input energy retained in the superbubbles ($\eta$) as a function of time for the different ambient densities. For all plots, the vertical axis denotes $\eta$. The coloured palette and the horizontal axis of the insets (a), (b) represent the dynamical time. This figure shows that $\eta$ decreases as the density of the ambient medium increases but asymptotically approaches
a constant value in time (see the inset (a)). The black solid line shows a rough scaling $\eta \propto \rho^{-1/3}_{amb}$ which matches well at almost all times for density $\rho_{amb}\gtrsim 100\, m{\rm_{H}\,cm^{-3}}$. The inset (b) displays $\eta$ in the presence/absence of different physical processes for the runs `SB\_d3\_ '. This figure shows that the dotted red curve (SB\_d3\_HCT) and the solid red curve (SB\_d3\_RHCT) coincide after $\gtrsim 3$ Myr (the time when the first supernova occurs, Figure \ref{fig:outSB99}).}
\label{fig:fracenergy}
\end{figure}
\begin{figure*}
\centering
\includegraphics[height=3.1in,width=6.5in]{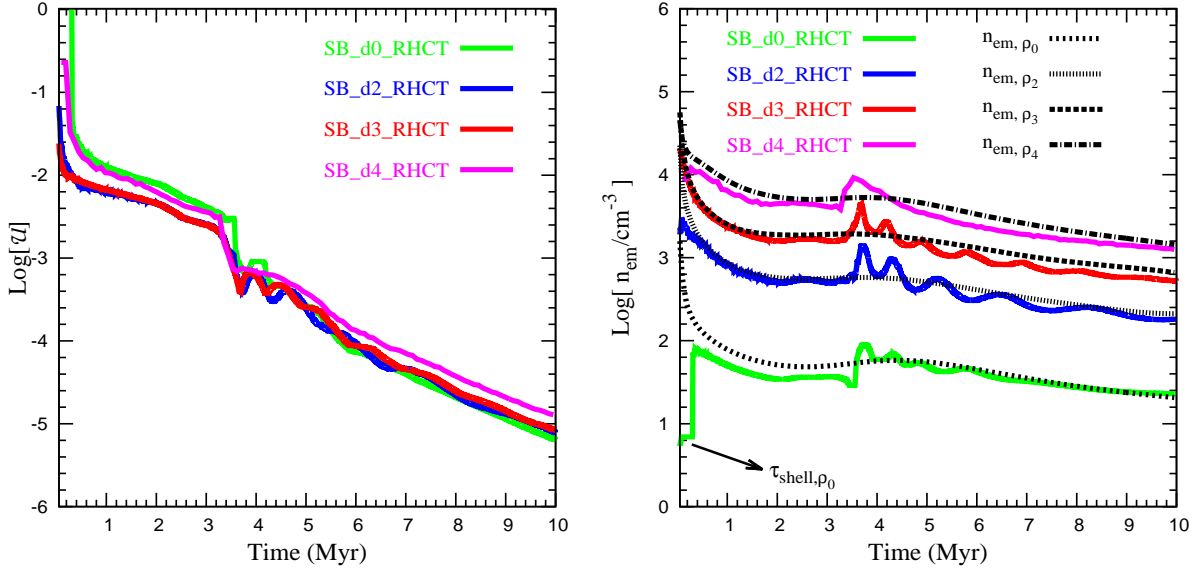} 
\caption{Observational parameters as a function of time for four different ambient density models. Left panel displays time evolution of ionization parameter $\mathcal{U}$. Time evolution of the recombination-averaged density $n_{em}$ is shown in the right panel. Here the symbol $\rho_{\rm x}$ represents the ambient density $10^{\rm x}\, m_{\rm H}$ cm$^{-3}$. A sudden drop of $\mathcal{U}$ (left panel) and hump in $n_{em}$ (right panel) at $\approx3.4$ Myr are due to the drop in ionizing photons flux and huge mass ejection because of SNe respectively (see Figure \ref{fig:outSB99}). The black lines in the right panel represent results obtained using $n_{em}\approx\mathcal{M}^2(T_{amb}/T_{i}\mu)\rho_{amb}/m{\rm_{H}}$, for details see equation (\ref{eq:dishell}).}
\label{fig:un}
\end{figure*}

To find the fraction of the retained input energy in superbubble, we have defined efficiency as
\begin{equation}
\label{eq:eta}
\eta= \frac{\Delta TE + KE}{E_{IN}}\, ,
\end{equation}
where $E_{IN}$ is the total amount of injected energy (i.e., work done by the radiation $E_{rad}$ and mechanical energy $E_{w}$) until a given epoch, $\Delta TE$ is the change in thermal energy (i.e., $\Delta TE = TE(t_{dyn}) - TE(t_{dyn}=0)$) and  $KE$ is the total kinetic energy in the simulation box at a given time (for details see Appendix \ref{app:techdetails}). Figure \ref{fig:fracenergy} displays $\eta$ as a function of dynamical time for four different ambient densities. The inset (a) shows that $\eta$ decreases monotonically as the density of the ambient medium increases, but asymptotically approaches a constant value in time. The inset (b) displays the time evolution of $\eta$ for the runs SB\_d3\_CT, SB\_d3\_HCT and SB\_d3\_RHCT. This figure shows that, for the runs SB\_d3\_HCT and SB\_d3\_RHCT, $\eta$ asymptotically approaches a nearly coincident value after $\gtrsim3$ Myr (corresponds to the epoch of the first supernova; see Figure \ref{fig:outSB99}).

From Figure \ref{fig:fracenergy}, we find that the asymptotic value of the energy efficiency  $\eta\approx0.1$ for $\rho_{amb}=10^{3}$ $m{\rm_{H}}$ cm$^{-3}$, and for $\rho_{amb}\gtrsim 100\, m{\rm_{H}\,cm^{-3}}$, the scaling of $\eta$ with ambient density is roughly $ \eta \propto \rho^{-1/3}_{amb}$. It is worth noting that low resolution simulations can show a lower efficiency due to over-cooling of unresolved regions (as highlighted by \citealt{Gentry2016}; \citealt{Yadav2016}). We note that the typical resolution (for the realistic runs) in our simulation is $\delta r =0.025-0.04$ pc (see Table \ref{table:model}) which is much higher than in typical 3-D simulations.

Note that the mechanical efficiency $\eta_{mech}$ introduced in equation (\ref{eq:ratioPrPsw}) is similar to the definition of $\eta$ (equation (\ref{eq:eta})), except that, for $\eta_{mech}$, $E_{IN}$ represents only the mechanical energy (i.e., $\eta_{mech}=(\Delta TE + KE)/E_{mech}$). However, after the end of radiation pressure dominating regime, $\eta_{mech}$ can be considered to be the same as $\eta$. Therefore, we may expect a similar scaling i.e., $\eta_{mech}\propto\rho^{-1/3}_{amb}$. In that case, using equation (\ref{eq:ratioPrPsw}), one can find that $\epsilon_{rad}\propto \rho^{1/15}_{amb}$. This also explains the result that the ratio ($\epsilon_{rad}$) of the radiation pressure ($P_{rad}$) to the thermal pressure ($P_{sw}$) is insensitive to the ambient densities for our realistic runs (Figure \ref{fig:ratioradprs}).

%
\subsection{Observational parameters}
\label{subsec:unem}
YM12 proposed various parameters to interpret observation and concluded that the ionization parameter $U$ can be used as a proxy to determine the dominant feedback mechanism for massive star clusters. The ionization parameter at the edge (i.e., $r\approx R_{cd}$) of the ionized shell is defined as 
\begin{equation}
\label{eq:U}
U = \frac{Q_{i}}{4\pi\, r^{2}\, n\, c}\,,
\end{equation}
which can be written as $U\sim$ $(k_{B}T_{i}/\left<h\nu\right>_{i})(P_{rad}/P_{\rm H{\scriptsize {II}}})$ and therefore, for a given $\left<h\nu\right>_{i}$ (i.e., for a stellar source of given radiation temperature), $U\propto P_{rad}/P_{\rm H{\scriptsize {II}}}$. Therefore, $U$ is directly connected to the observables. However, it is worth noting that for a realistic cluster, $\left<h\nu\right>_{i}$ is a function of time.

The above definition of ionization parameter is useful when the density of the ionized medium is uniform. For a non-uniform density, YM12 suggested an expression 
\begin{equation}
\label{eq:Uem}
\mathcal{U} = \frac{\int U(r)\,n^{2}\,dV }{\int n^{2} dV}\ ,
\end{equation}
where the integration begins at the inner edge ($r=R_{cd}$) of ionized shell and ends at the outer edge of ionized shell, $dV$ is the elementary volume which is equal to $4\pi r^{2} dr$ (1-D spherical). The recombination-averaged density is given by
\begin{equation}
\label{eq:nem}
n_{em} = \frac{\int n\,n^{2} dV }{\int n^{2} dV}
\end{equation}
here the limit of the integration is same as in equation (\ref{eq:Uem}). Observationally, one can compute the value of $\mathcal{U}$ and $n_{em}$ by comparing the strength of different spectral lines (for details see YM12).

\citealt{Martinez2014} (hereafter MST14) have estimated $\mathcal{U}$ and found that $\mathcal{U}$ is almost constant in time but depends on the density of the ambient medium in the absence of the SW region (i.e.; radiation dominated regime in their case; see Figure $8$ in MST14). Note that, they used equation (\ref{eq:tcoolsw}) (\citealt{MacLow1988}) to find the cooling time scale of the SW region. However, we find that in the presence of radiative cooling, SW region is always present for a realistic source parameters (see model SB\_d3\_CT in Figure \ref{fig:drprsSW}). A more {\it realistic} time evolution of $\mathcal{U}$ for four different ambient densities are shown in the left panel of Figure \ref{fig:un}.

The right panel of Figure \ref{fig:un} displays the recombination-averaged density ($n_{em}$) as a function of time for four different ambient densities. The computed values of $n_{em}$ match well with the estimates of $n_{em}\approx \rho_{i\_shell}/(\mu m{\rm_H})$ from equation (\ref{eq:dishell}) (shown by black curves). This match provides a method of estimating the Mach number from observations of the strength of the spectral lines, which is related to $n_{em}$ (YM12).

It is worth noting that, although $n_{em}$ depends on $\rho_{amb}$, $\mathcal{U}$ is not sensitive to $\rho_{amb}$. This is because, for a given $Q_{i}$, $U\propto(1/n\,r^{2})$ and in presence of radiative energy loss, high density medium suffers more radiative energy loss compare to low density medium which makes $n\,r^{2}$ almost independent of the ambient density (see Figure \ref{fig:ratioradprs}, which shows that $P_{rad}/P_{sw}$ falls within the same range for $1\lesssim\rho_{amb}\lesssim10^{4}$ $m\rm{_{H}\,cm^{-3}}$). Also note that, at early times ($\lesssim3$ Myr), $-1.6\lesssim$ Log[$\mathcal{U}$] $\lesssim- 2.5$ which is consistent with observed value of $\mathcal{U}$ for starburst galaxies (e.g., Log[$\mathcal{U}$]$\approx - 2.3$ for M82, NGC3256 and NGC 253, see Table $4$ in YM$12$).
\subsection{Application to 30 Doradus}
\label{subsec:othermodlcomp} 
\citealt{Lopez2011} and \citealt{Pellegrini2011} have interpreted the observations of 30 Doradus differently and reached a somewhat different conclusions with regard to its dynamics. 

\citealt{Lopez2011} estimated the radiation pressure at a distance $r$ due to individual star and then taking a sum over all stars, they have defined $P_{dir}=\sum(L_{bol}/4\pi r^2 c)$. They compared $P_{dir}$ with the thermal pressure of X-ray plasma $P_{X}$ (which is equivalent to the comparison of $P_{rad}$ with $P_{sw}$) and found that $P_{dir}\gtrsim P_{X}$ when $r \lesssim 75$ pc. From this, they argued that the expansion of 30 Doradus at early time is in radiation dominated regime (for details see section 3 and 5 in their paper). Approaching the problem in a different way, \citealt{Pellegrini2011}  have estimated the ionization parameter of the photoionized region of the shell and defined the radiation pressure at distance $r$ due to absorption of incident starlight $P_{star} = U n_{H}\left<h\nu\right> L_{bol}/L_{i}$, where $\left<h\nu\right>$ ($\approx 20$ eV) is the average energy per photon. From this, they have shown that the ratio $P_{star}/P_{gas}$ drops below $1/3$ when $r\gtrsim10$ pc and concluded that radiation has negligible importance in the dynamics of 30 Doradus (for details see section 3 in their paper).

ST13 have shown that, in a high density medium ($\sim 10^{3}$ cm$^{-3}$), $P_{rad}/P_{sw}$ exceeds unity only after the bubble makes the transition from energy dominated regime to momentum dominated regime (i.e., in the absence of the SW region) and concluded that radiation pressure is unlikely to control the dynamics of 30 Doradus. MST14 took one special case (HDE: High-density with low heating efficiency) where they used the same $Q_{i}$ and $L_{i},L_{n}$ with one order magnitude less $L_{w}$ but even in that case they found that the role of radiation pressure is important {\it after} $\sim0.85$ Myr. On the contrary, using realistic simulations, we have found that radiation pressure controls the dynamics at early time $\lesssim 1$ Myr. As time evolves, the strength of radiation pressure decreases because of $1/r^{2}$ dependence and also due to sudden fall of $L_{i}$ after $3.4$ Myr. Therefore, we find that \citealt{Lopez2011} over-estimated the role of radiation and \citealt{Pellegrini2011} under-estimated it. However, our result is consistent with one aspect that at early times the dynamics of 30 Doradus is controlled by radiation pressure. 
\section{Summary}
\label{sec:summary}
In this paper, we have focused on the effects of winds and radiation on the dynamics of supperbubbles in dense medium ($\rho_{amb}\gtrsim10^{2}$ $m{\rm_{H}\, cm^{-3}}$). We have performed high resolution $1$-D simulations and used a realistic time evolution of the mechanical and radiation power of a young star cluster of mass $10^{6}$ M$_{\odot}$. We stress on the importance of radiative cooling and heating in bubble evolution. We have explored the parameter space of the ambient density and dust absorption coefficient. We have calculated the temperature distribution of cooling losses and the energy efficiency of the superbubbles, discussed the observational parameters and compared our results with the observations of 30 Doradus. Our main results are summarised as follows

\begin{enumerate}
\item {\it Structure of a realistic ISB } : In the presence of radiative cooling, for a given mechanical luminosity, the internal structure of ISB depends on the mass-loss rate of the driving source. For high mass-loss rate, ISB can take longer time to form the SW region. But, once the SW region is formed, its cooling time scale becomes longer than dynamical time and it does not disappear (Figure \ref{fig:coolzoom}). 
\item {\it The effective dynamical force} : The ratio of radiation pressure to thermal pressure in the SW region is greater than unity before $\lesssim1$ Myr (Figure \ref{fig:ratioradprs}). This conclusion remains same when the density of the ambient medium $1\lesssim\rho_{amb}\lesssim10^{4}\ m\rm{_{H}\,cm^{-3}}$, but it may depend on the evolutionary profile of the input source. At an early time, the radiation pressure may play an important role in launching the shock (Figure \ref{fig:comprzoomin}), as a consequence it can affect star formation within the cluster volume itself. However, its strength decreases with time, because $P_{rad}\propto1/r^{2}$ and also because of the rapid decrease of the radiation luminosity after the first supernova. At a later time, the dynamics of the bubble is controlled by radiation heating and by thermal pressure of the SW, rather than the radiation pressure.
\item {\it Dust opacity dependence} : For a given ambient density and input source profile, the size of the central cavity depends weakly on the dust opacity ($\sigma_{d} \approx(0.1-1.5)\times 10^{-21}$ cm$^{2}$) of the ambient medium. However, the structure of shell depends on $\sigma_{d}$ (Figure \ref{fig:diffk}). A lower value of $\sigma_{d}$ enhances the heating efficiency of the input radiation field, and hence, increases the width of the photoionized region within the shell.
\item {\it Cooling losses \& the retained energy} : Most of the radiative losses occur at $\sim10^{4}$ K, with sub-dominant losses at $\lesssim 10^{3}$ K and $\sim10^{6}-10^{8}$ K (Figure \ref{fig:cooltempdis}). For $\rho_{amb}\gtrsim 10^{2}$ $m{\rm_{H}}$ cm$^{-3}$, the 
scaling of $\eta$ (fraction of the retained input energy in superbubble) with ambient density ($\rho_{amb}$) is roughly $ \eta \propto \rho^{-1/3}_{amb}$. The asymptotic value of $\eta$ is $\approx0.1$ for $\rho_{amb}=10^{3}$ $m{\rm_{H}}$ cm$^{-3}$(Figure \ref{fig:fracenergy}).

\item {\it Observational parameters} : The ionization parameter is weakly sensitive to the ambient density (Figure \ref{fig:un}). The recombination-averaged density ($n_{em}$) depends on the velocity of the expanding shell. If some independent estimate of the shell velocity and ambient density is available, then equation (\ref{eq:dishell}) can be used to predict the density of the ionized shell.
\end{enumerate}
\section*{Acknowledgements}
We thank Sergiy Silich and Sergio Mart\'{i}nez-Gonz\'{a}lez for providing the set-up details of Starburst99. We thank Naveen Yadav for comparing the mechanical energy efficiency between $1$-D and $3$-D models. SG thanks Arpita Roy, Deovrat Prasad and Kartick C. Sarkar for helpful discussions. Research of SG is supported by SPM Fellowship from CSIR, India. PS acknowledges the Department of Science and Technology, India grant no. Sr/S2/HEP-048/2012 and an India-Israel joint research grant (6-10/2014[IC]). YS acknowledges 15-52-45114-IND from RFBR.
\footnotesize{

\clearpage
\appendix
\section{Ionization state of a gas}
\label{app:cool-sim}
\begin{figure}
\centering
\includegraphics[height=2.25in,width=2.7in]{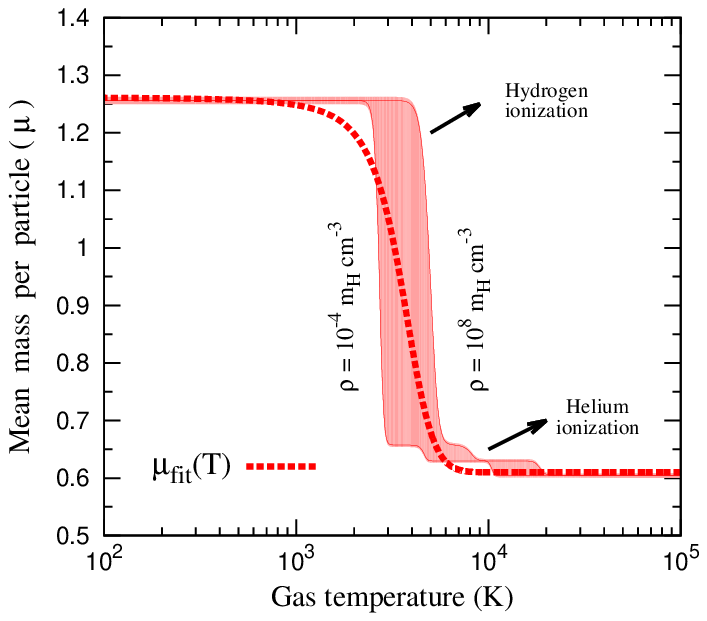} 
\caption{Dependence of the mean mass per particle $\mu$ (normalised w.r.t $m_{\rm H}$) on the gas temperature. The shaded region displays the temperature range between the ionized state and neutral state, which is obtained by considering the ionization of Hydrogen and Helium, and by solving Saha equation for different gas densities. The red-dashed line is a fitted function $\mu_{fit}(T)=0.61\left[1+1.075/(1+2\times e^{(T-4000)/700})\right]$.}
 \label{fig:varimu}
\end{figure}
The state of ideal gas is mainly characterised by three variables : pressure ($P$), density ($\rho$) and temperature ($T$). At a given time, Euler equations can obtain the solutions for two of them, the third variable temperature is directly calculated from relation 
\begin{equation}
T = \frac{\mu\,m_{H}}{k_{B}}\frac{P}{\rho}\ ,
\end{equation}
where $\mu$ is the mean mass per particle (normalised w.r.t $m_{\rm H}$). The numerical value of $\mu$ depends on the gas composition and the ionization state of the gas. For a completely neutral pure hydrogen gas $\mu$ is $1$ and for completely ionized gas $\mu=0.5$. In our case, we assume metallicity of the gas $Z\approx0.4 Z_{\odot}$ and therefore, for neutral ISM $\mu\approx1.26$, and for completely ionized gas $\mu\approx 0.61$ and the mean mass per ion $\mu_{i}\approx1.275$.

We have assumed the initial temperature of the ambient medium to be $T_{amb}=100$ K. At this temperature, chemistry is important and  one should do a full analysis of ionization fraction of each species. But for practical purposes, we take this into consideration by using a fitting function for $\mu$ which is shown in Figure \ref{fig:varimu}. At each step, the calculation starts with $\mu=1.0$ to estimate a dummy $T$ and then it uses the fit function to minimise the error between consecutive $T<0.1\%$. Once it able to find $T$, it calculates mean mass per electron ($\mu_{e}$=1/$\left[1/\mu-1/\mu_{i}\right]$) and also $\Lambda_{N}$($T,Z$) from the tabulated cooling curve listed in PLUTO.

\section{Luminosity fraction}
\label{app:phi}
When the ionizing and non-ionizing photons travel through a dusty medium, a significant fraction of them are absorbed by the dust. The attenuation fractions of the ionizing and non-ionizing luminosity are determined by (see equations $2$ and $3$ in \citealt{Draine2011b})
\begin{equation}
\label{eq:phiI}
\frac{d\phi_{i}}{dr} = - \frac{4 \pi\,\alpha_{B}}{Q_{i}}\,n^{2}\, x^{2}\, r^2 - n \sigma_{d}\phi_{i}
\end{equation}
\begin{equation}
\label{eq:phiNI}
\frac{d\phi_{n}}{dr} =  - n \sigma_{d}\,\phi_{n}
\end{equation}
\begin{figure}
\centering
\includegraphics[height=2.25in,width=2.7in]{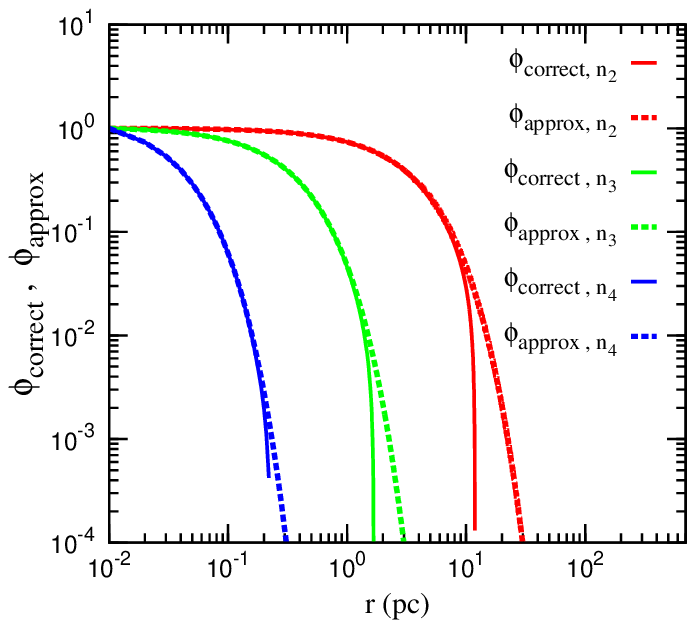} 
\caption{Comparison of luminosity fractions ($\phi$) as a function of the distance $r$ for three different densities ($n_{\rm x}=10^{\rm x}$ cm$^{-3}$). Here the distance $r$ is measured from the inner edge of the H{\scriptsize {II}} region i.e., $r\equiv r-R_{cd}$, $R_{cd}$ is the position of contact discontinuity. This figure shows that $\phi_{\rm approx}$ is almost same as $\phi_{\rm correct}$.}
 \label{fig:phi}
\end{figure}
where $\alpha_{B}$ is `Case B' recombination coefficient, $x$ is the ionization fraction of hydrogen (i.e, for completely ionized medium $x=1$; for details see section \ref{subsec:heatset}) and the distance $r$ is measured from the inner edge of the H{\scriptsize {II}} region. The first term in equation (\ref{eq:phiI}) represents the loss of the ionizing photons due to photoionization of hydrogen and second term represents dust absorption.

In equation (\ref{eq:phiI}), if first term is small compare to second term, then both equations have same solution $\phi_{i}\approx\phi_{n}=e^{-\tau}$, where $\tau=\int{n\,\sigma_{d}\,dr}$ ($\sigma_{d}$ is the dust absorption coefficient). Assuming $x=1$, the solution of equations (\ref{eq:phiI}) and (\ref{eq:phiNI}) for three different uniform densities are shown in Figure \ref{fig:phi}. This figure shows that for high density medium the difference between $\phi_{i}\,(=\phi_{\rm correct})$ and $\phi_{n}\,(=\phi_{\rm approx})$ is small (except at the edge where $\phi_{i}$ drops faster than $\phi_{n}$). Note that for a realistic case, the shell density is not uniform, at the outer edge of the H{\scriptsize {II}} region $x\neq 1$ and because of this, the choice $\phi_{i}\approx\phi_{n}=\,e^{-\tau}$ is more robust compare to the actual solution.
\section{Conservation test}
\label{app:techdetails}
Conservation test is essential for any simulation set-up. To check this, we have defined simulation energy ($E$) efficiency as 
\begin{equation}
\epsilon_{E}(t) = \frac{TE(t) + KE(t) + E_{eff, RL}(t)}{TE(t=0) + E_{IN}(t)}
\end{equation} 
where $TE$ and $KE$ are the total thermal energy and kinetic energy in the simulation box at a given time ($t$), $E_{eff, RL}$ is the effective energy loss due to radiative cooling (plus heating) until a given epoch (i.e, $E_{eff, RL}=-|E_{q^{-}}|+E_{q^{+}}$, where  $E_{q^{-}}$ and $E_{q^{+}}$ represent the terms associated with radiative cooling and heating respectively). $E_{IN}$ is the sum of total mechanical energy ($E_{w}(t)=\int dt\, L_{w}$) and radiation energy ($E_{rad}(t)=\int_{{\small t}} \int_{{\small r}}\, dt\, dr\ 4\,\pi\,r^{2}\ \vec{v}.(\rho\,\vec{a}_{rad}$), where $\rho\,a_{rad}$ is radiation force per unit volume, see equation (\ref{eq:radsrc})) until a given epoch. 

The mass ($M$) efficiency is defined as
\begin{equation}
\epsilon_{M}(t) = \frac{M_{box}(t)}{ M_{box}(t=0) + M_{IN}(t)},
\end{equation}
where $M_{box}$ is the total mass in the simulation box at a given time $t$ and $M_{IN}$ is the total added mass until that time.

Therefore, according to above definitions, if $\epsilon_{X} = 1.0$, then the quantity $X$ is conserved. Figure \ref{fig:figconrv} displays $\epsilon_{X}$ as a function of time for all of the runs (see Table \ref{table:model}) and confirms that conservation holds with accuracy $\gtrsim99.5\,\%$.
\begin{figure}
\centering
\includegraphics[height=2.25in,width=3.25in]{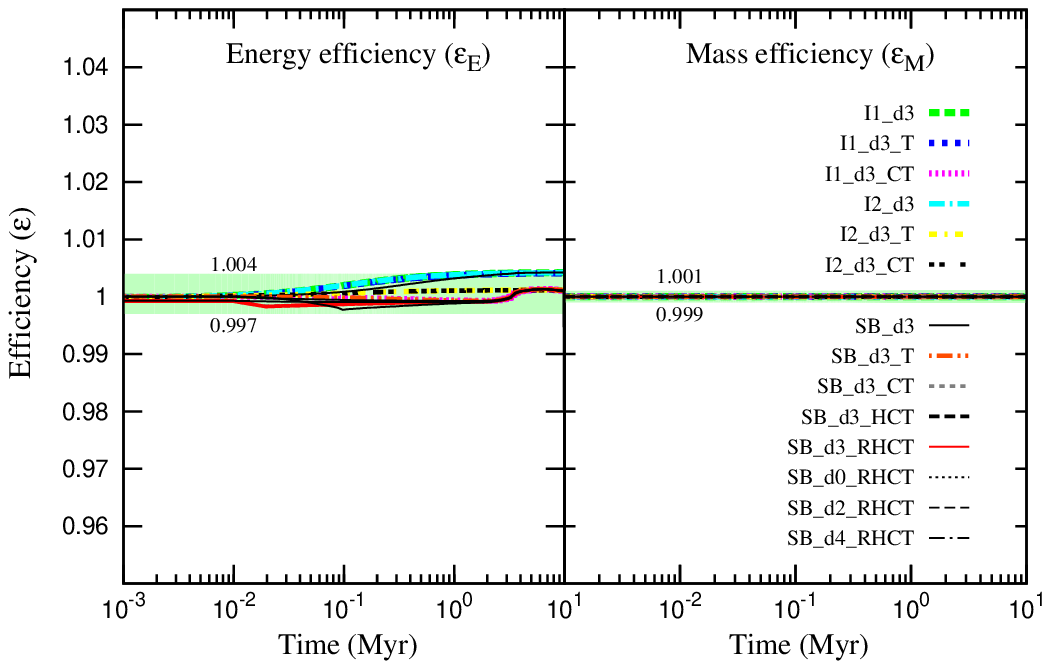} 
\caption{Efficiency of energy and mass as a function of time for all runs. Left plot (a) shows that maximum error in energy is $\lesssim0.4\%$ which indicates accuracy in energy budget $\gtrsim$ $99.5\%$. Right figure (b) shows that accuracy in mass budget is more than $99.9\%$.}
 \label{fig:figconrv}
\end{figure}

\section{Convergence Test}
\label{app:resolution}
In addition to conservation test, we have done the resolution test for all the models and find that our conclusions remain same. Here we have shown one particular model that corresponds to the formation of the SW region.

Figure \ref{fig:app_coolzoom} displays cooling time scale of the SW region (or pre-shocked wind layer) for the model I1\_d3\_CT as a function of time for three different resolutions ($\delta r$). This figure shows that cooling time scale of pre-shocked wind layer is shorter than $t_{dyn}$ at early times, and it becomes longer than $t_{dyn}$ after $\gtrsim 0.3$ Myr. When $\tau_{sw}$ is longer than $t_{dyn}$ ($\gtrsim 0.3$ Myr) the SW region is formed. Before $\sim 0.3$ Myr, there are few spikes which are due to the rapid cooling of the pre-shock wind layer. Note that, for the {\it realistic} models (model label : SB\_, Table \ref{table:model}), we have used output of Starburst99 which do not show these spikes (see blue curve in Figure \ref{fig:coolzoom}). Therefore, the spikes shown here highlight a special case for unrealistically large mass-loss rate, showing the formation of the SW region in the presence of excessive radiative cooling.
\begin{figure}
\centering
\includegraphics[height=2.25in,width=2.7in]{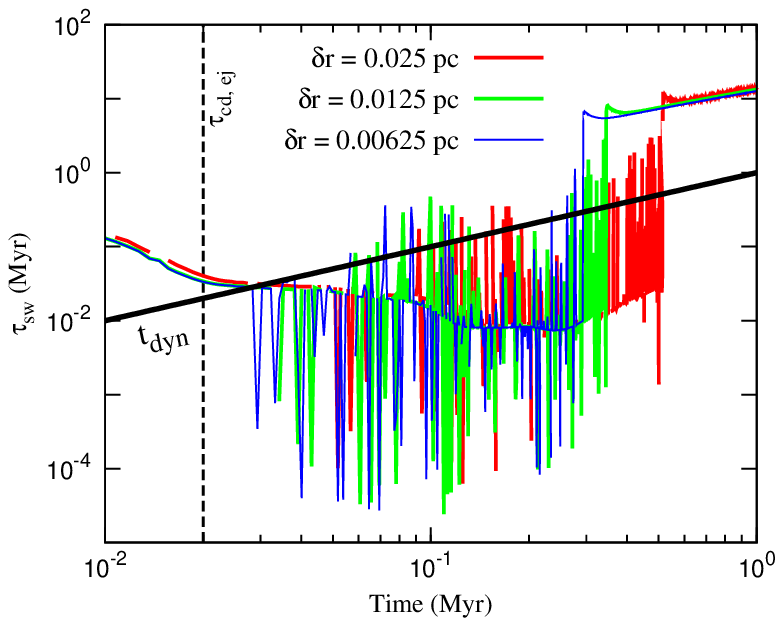} 
\caption{Cooling time scale of the pre-shocked wind layer or SW region ($\tau_{sw}$) as a function of time for three different resolutions $\delta r = 0.025$ pc, $0.0125$ pc and $0.00625$ pc for the run I1\_d3\_CT (large mass-loss rate). Solid black line represents the dynamical time ($t_{dyn}$). The vertical dashed line represents cooling time scale of the pre-shocked wind layer $\tau_{cd,ej}$ using equation (\ref{eq:tcoolej}). At early times, the sawtooth-like behaviour of the curves are connected with thermal instability because of the excessive cooling of the pre-shock wind layer (also see Figure \ref{fig:coolzoom} which displays $\tau_{sw}$ for the resolution $\delta r=0.025$ pc).}
 \label{fig:app_coolzoom}
\end{figure}
\section{Dependence on cluster mass}
\label{app:Mcluster}
\begin{figure}
\centering
\includegraphics[height=2.25in,width=2.9in]{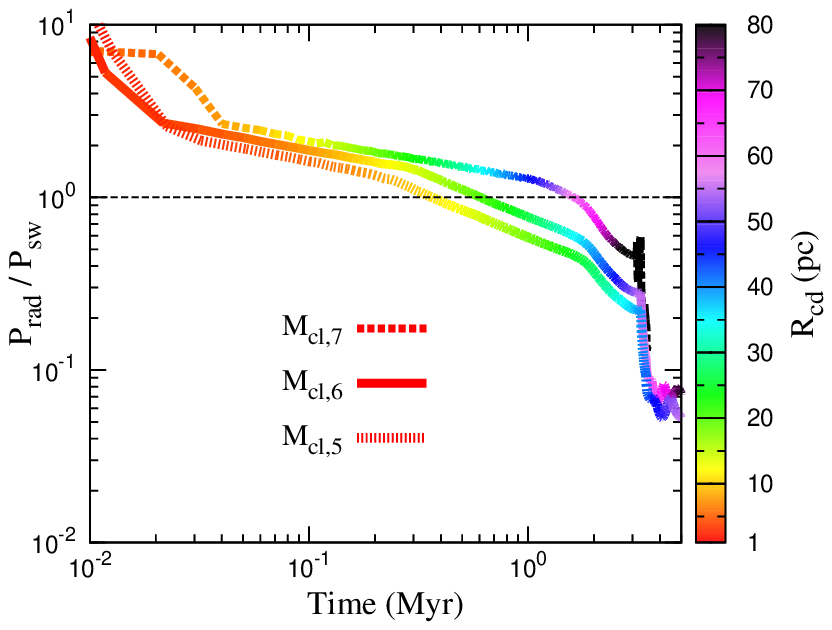} 
\caption{Evolution of the ratio of radiation pressure ($P_{rad}$) to thermal pressure of the SW region ($P_{sw}$) for three different masses of star cluster. The symbol ${\rm M_{cl, x}} = 10^{\rm x} \,{\rm M_{\odot}}$ denotes the mass of the star cluster. For all cases, the ambient density is taken as $10^{3}\,m\rm{_{H}\,cm^{-3}}$ and represents a realistic run (i.e., SB\_d3\_RHCT; see Table \ref{table:model}). The colour palette represents the position of the contact discontinuity ($R_{cd}$). This figure shows that the ratio $P_{rad}$ to $P_{sw}$ slowly increases with $M_{cl}$. For $M_{cl}=10^{6}\,{\rm M_{\odot}}$, see Figure \ref{fig:ratioradprs}.}
 \label{fig:PraddiffM}
\end{figure}
Throughout this paper, we have considered a star cluster of mass $10^{6}\,{\rm M_{\odot}}$, and we concluded that $\epsilon_{rad}=P_{rad}/P_{sw}$ is greater than unity before $1$ Myr. Here we discuss the dependence of $\epsilon_{rad}$ on the mass of the star cluster ($M_{cl}$).
As already shown in section \ref{subsec:radiation}, $\epsilon_{rad}$ is insensitive to the ambient density, and at a given epoch, $\epsilon_{rad}$ is roughly proportional to $L_{bol}^{1/5}$. For a given Initial Mass Function, $L_{bol}\propto M_{cl}$ (\citealt{Leitherer99}), therefore $\epsilon_{rad}\propto M_{cl}^{1/5}$. Figure \ref{fig:PraddiffM} displays the ratio of $P_{rad}$ to $P_{sw}$ as a function of time for three different masses of star cluster ($M_{cl} =$ $10^{7}$, $10^{6}$ and $10^{5}\,{\rm M_{\odot}}$). This figure shows that the role of radiation pressure is important for massive star clusters.

\begin{thebibliography}{}
\bibitem[\protect\citeauthoryear{Alexiades \etal}{1996}]{Alexiades1996}
Alexiades, V., Amiez, G., \& Gremaud P.-A. 1996,
 \ufhref[webgreen]{http://dx.doi.org/10.1002/(SICI)1099-0887(199601)12:1<31::AID-CNM950>3.0.CO;2-5}{Com. Num. Meth. Eng.},
 \ufhref[webgreen]{http://dx.doi.org/10.1002/(SICI)1099-0887(199601)12:1<31::AID-CNM950>3.0.CO;2-5}{12, 31}
 
\bibitem[\protect\citeauthoryear{Bisbas \etal} {2015}]{Bisbas2015}
Bisbas, T. G., Haworth, T. J., Williams, R. J. R., Mackey, J., Tremblin, P. \etal \ 2015
 \ufhref[webgreen]{http://dx.doi.org/10.1093/mnras/stv1659}{MNRAS}, 
 \ufhref[webgreen]{http://adsabs.harvard.edu/abs/2015MNRAS.453.1324B}{453, 1324} 
 
 
 \bibitem[\protect\citeauthoryear{Bisnovatyi-Kogan \& Silich} {1995}]{Bisnovatyi1995}
Bisnovatyi-Kogan, G. S., \& Silich, S. A. 1995,
 \ufhref[webgreen]{http://adsabs.harvard.edu/abs/1995RvMP...67..661B}{RvMP}, 
 \ufhref[webgreen]{http://adsabs.harvard.edu/abs/1995RvMP...67..661B}{ 67, 661} 

 
\bibitem[\protect\citeauthoryear{Castor \etal} {1975}]{Castor1975}
Castor, J., McCray, R., Weaver, R. 1975
 \ufhref[webgreen]{http://dx.doi.org/10.1086/181908}{ApJ}, 
 \ufhref[webgreen]{http://adsabs.harvard.edu/abs/1975ApJ...200L.107C}{200, 107}

\bibitem[\protect\citeauthoryear{Chevalier \& Clegg}{1985}]{CC1985}
Chevalier, R. A. \& Clegg, A. W. 1985,
 \ufhref[webgreen]{http://dx.doi.org/10.1038/317044a0}{Nature}, 
 \ufhref[webgreen]{http://adsabs.harvard.edu/abs/1985Natur.317...44C}{317, 44}

\bibitem[\protect\citeauthoryear{Cowie \& McKee}{1977}]{Cowie1977}
Cowie, L. L., \& McKee, C. F. 1977,
 \ufhref[webgreen]{http://dx.doi.org/10.1086/154911}{ApJ}, 
 \ufhref[webgreen]{http://adsabs.harvard.edu/abs/1977ApJ...211..135C}{221, 135}

\bibitem[\protect\citeauthoryear{Dale} {2015}]{Dale2015}
Dale, J. E. 2015,
 \ufhref[webgreen]{http://dx.doi.org/10.1016/j.newar.2015.06.001}{NewAR}, 
 \ufhref[webgreen]{http://adsabs.harvard.edu/abs/2015NewAR..68....1D}{68, 1}

\bibitem[\protect\citeauthoryear{Dale \etal} {2013}]{Dale2013}
Dale, J. E., Ercolano, B., Bonnell, I. A. 2013,
 \ufhref[webgreen]{http://dx.doi.org/10.1093/mnras/sts592}{MNRAS}, 
 \ufhref[webgreen]{http://adsabs.harvard.edu/abs/2013MNRAS.430..234D}{430, 234}
 
\bibitem[\protect\citeauthoryear{Draine} {2011a}]{Draine2011a}
Draine, B. T. 2011a, Physics of the Interstellar and Intergalactic Medium,
 \ufhref[webgreen]{http://adsabs.harvard.edu/abs/2011piim.book.....D}{(Princeton University Press)}

\bibitem[\protect\citeauthoryear{Draine} {2011b}]{Draine2011b}
Draine, B. T. 2011b,
 \ufhref[webgreen]{http://dx.doi.org/10.1088/0004-637X/732/2/100}{ApJ}, 
 \ufhref[webgreen]{http://adsabs.harvard.edu/abs/2011ApJ...732..100D}{732, 100}



\bibitem[\protect\citeauthoryear{Federrath} {2015}]{Federrath2015}
Federrath, C. 2015,
 \ufhref[webgreen]{http://dx.doi.org/10.1093/mnras/stv941}{MNRAS}, 
 \ufhref[webgreen]{http://adsabs.harvard.edu/abs/2015MNRAS.450.4035F}{450, 4035}
 
\bibitem[\protect\citeauthoryear{Ferland \etal} {1998}]{Ferland1998}
Ferland, G. J., Korista, K. T., Verner, D. A., Ferguson, J. W., Kingdon, J. B., \& 
Verner, E. M. 1998,
 \ufhref[webgreen]{http://dx.doi.org/10.1086/316190}{PASP}, 
 \ufhref[webgreen]{http://adsabs.harvard.edu/abs/1998PASP..110..761F}{110, 761}

\bibitem[\protect\citeauthoryear{Ferrara \& Shchekinov} {1993}]{Ferrara1993}
Ferrara, A \& Shchekinov, Y. 1993,
 \ufhref[webgreen]{http://dx.doi.org/10.1086/173339}{ApJ}, 
 \ufhref[webgreen]{http://adsabs.harvard.edu/abs/1993ApJ...417..595F}{417, 595}


\bibitem[\protect\citeauthoryear{Geen \etal} {2015}]{Geen2015}
Geen, S., Rosdahl, J., Blaizot, J., Devriendt, J., Slyz, A. 2015,
 \ufhref[webgreen]{http://dx.doi.org/10.1093/mnras/stv251}{MNRAS}, 
 \ufhref[webgreen]{http://adsabs.harvard.edu/abs/2015MNRAS.448.3248G}{448, 3248}

\bibitem[\protect\citeauthoryear{Gentry \etal} {2016}]{Gentry2016}
Gentry, E. S., Krumholz, M. R., Dekel, A. \& Madau, P. 2016,
 \ufhref[webgreen]{http://arxiv.org/abs/1606.01242}{arXiv:1606.01242}


\bibitem[\protect\citeauthoryear{Harper-Clark \& Murray} {2009}]{Harper2009}
Harper-Clark, E. \& Murray, N. 2009, 
 \ufhref[webgreen]{http://dx.doi.org/10.1088/0004-637X/693/2/1696}{ApJ}, 
 \ufhref[webgreen]{http://dx.doi.org/10.1088/0004-637X/693/2/1696}{693, 1696}

\bibitem[\protect\citeauthoryear{Haworth \etal} {2015}]{Haworth2015}
Haworth, T. J., Harries, T. J., Acreman, D. M., Bisbas, T. G. 2015,
 \ufhref[webgreen]{http://dx.doi.org/10.1093/mnras/stv1814}{MNRAS}, 
 \ufhref[webgreen]{http://adsabs.harvard.edu/abs/2015MNRAS.453.2277H}{453, 2277}
 
\bibitem[\protect\citeauthoryear{Hopkins \etal} {2011}]{Hopkins2011}
Hopkins, P. H., Quataert, E \& Murray, N. 2011,
 \ufhref[webgreen]{doi:10.1111/j.1365-2966.2011.19306.x}{MNRAS}, 
 \ufhref[webgreen]{http://arxiv.org/abs/1101.4940}{417, 950}


\bibitem[\protect\citeauthoryear{Kim \etal}{2016}]{Kim2016} 
Kim, J., Kim, W., \& Ostriker, E. C. 2016,
 \ufhref[webgreen]{http://arxiv.org/abs/1601.03035}{ApJ in press}, 
 \ufhref[webgreen]{http://arxiv.org/abs/1601.03035}{arXiv:1601.03035}
 
\bibitem[\protect\citeauthoryear{Kroupa} {2002}]{Kroupa2002}
Kroupa, P., 2002,
 \ufhref[webgreen]{http://dx.doi.org/10.1126/science.1067524}{Science}, 
 \ufhref[webgreen]{http://adsabs.harvard.edu/abs/2002Sci...295...82K}{295, 82}
 
\bibitem[\protect\citeauthoryear{Krumholz \& Matzner} {2009}]{Krumholz2009}
Krumholz, M. R., \& Matzner, C. D. 2009,
 \ufhref[webgreen]{http://dx.doi.org/10.1088/0004-637X/703/2/1352}{ApJ}, 
 \ufhref[webgreen]{http://adsabs.harvard.edu/abs/2009ApJ...703.1352K}{703, 1352}
  
\bibitem[\protect\citeauthoryear{Leitherer \etal}{1999}]{Leitherer99}
Leitherer, C., Schaerer, D., Goldader, J. D., et al. 1999,
 \ufhref[webgreen]{http://dx.doi.org/10.1086/313233}{ApJS},
 \ufhref[webgreen]{http://adsabs.harvard.edu/abs/1999ApJS..123....3L}{123, 3}

\bibitem[\protect\citeauthoryear{Lopez \etal}{2011}]{Lopez2011}
Lopez, L. A., Krumholz, M. R., Bolatto, A. D., Prochaska, J. X., \& Ramirez-Ruiz, E. 2011,
 \ufhref[webgreen]{http://dx.doi.org/10.1088/0004-637X/731/2/91}{ApJ},
 \ufhref[webgreen]{http://adsabs.harvard.edu/abs/2011ApJ...731...91L}{731, 91}

\bibitem[\protect\citeauthoryear{Mac Low \& McCray}{1988}]{MacLow1988}
Mac Low, M.-M., \& McCray, R. 1988,
 \ufhref[webgreen]{http://dx.doi.org/10.1086/165936}{ApJ}, 
 \ufhref[webgreen]{http://adsabs.harvard.edu/abs/1988ApJ...324..776M}{324, 776}

\bibitem[\protect\citeauthoryear{Mart\'{i}nez-Gonz\'{a}lez \etal}{2014}]{Martinez2014}
Mart\'{i}nez-Gonz\'{a}lez, S., Silich S. \& Tenorio-Tagle G., 2014,
 \ufhref[webgreen]{http://dx.doi.org/10.1088/0004-637X/785/2/164}{ApJ}, 
 \ufhref[webgreen]{http://adsabs.harvard.edu/abs/2014ApJ...785..164M}{785, 164}

\bibitem[\protect\citeauthoryear{Mignone \etal} {2007}]{Mignone2007}
Mignone, A., Bodo, G., Massaglia, S., Matsakos, T., Tesileanu, O., Zanni, C., Ferrari, A., 2007,
 \ufhref[webgreen]{http://dx.doi.org/10.1086/513316}{ApJS}, 
 \ufhref[webgreen]{http://adsabs.harvard.edu/abs/2007ApJS..170..228M}{170, 228}

\bibitem[\protect\citeauthoryear{Murray \etal}{2011}]{Murray2011}
Murray, N., M\'{e}nard, B. \& Thompson, T. A. 2011,
 \ufhref[webgreen]{http://dx.doi.org/10.1088/0004-637X/735/1/66}{ApJ}, 
 \ufhref[webgreen]{http://adsabs.harvard.edu/abs/2011ApJ...735...66M}{735, 66}

 
\bibitem[\protect\citeauthoryear{Parker}{1965}]{Parker1965}
Parker, E. N. 1965,
 \ufhref[webgreen]{http://dx.doi.org/10.1007/BF00216273}{SSRv},
 \ufhref[webgreen]{http://adsabs.harvard.edu/abs/1965SSRv....4..666P}{4, 666}

\bibitem[\protect\citeauthoryear{Pellegrini \etal}{2011}]{Pellegrini2011}
Pellegrini, E. W., Baldwin, J. A., \& Ferland, G. J. 2011,
 \ufhref[webgreen]{http://dx.doi.org/10.1088/0004-637X/738/1/34}{ApJ},
 \ufhref[webgreen]{http://adsabs.harvard.edu/abs/2011ApJ...738...34P}{738, 34}

\bibitem[\protect\citeauthoryear{Rogers \& Pittard}{2013}]{Rogers2013}
Rogers, H., Pittard, J. M. 2013
 \ufhref[webgreen]{http://dx.doi.org/10.1093/mnras/stt255}{MNRAS},
 \ufhref[webgreen]{http://adsabs.harvard.edu/abs/2013MNRAS.431.1337R}{431, 1337}

\bibitem[\protect\citeauthoryear{Sales \etal}{2014}]{Sales2014}
Sales, L. V., Marinacci, F., Springel, V., Petkova, M. 2014
 \ufhref[webgreen]{http://dx.doi.org/10.1093/mnras/stu155}{MNRAS}, 
 \ufhref[webgreen]{http://adsabs.harvard.edu/abs/2014MNRAS.439.2990S}{439, 2990}

\bibitem[\protect\citeauthoryear{Sharma \etal}{2014}]{Sharma2014}
Sharma, P., Roy, A., Nath, B. B., Shchekinov, Y., 2014,
 \ufhref[webgreen]{http://dx.doi.org/10.1093/mnras/stu1307}{MNRAS}, 
 \ufhref[webgreen]{http://adsabs.harvard.edu/abs/2014MNRAS.443.3463S}{443, 3463}
 
\bibitem[\protect\citeauthoryear{Shu }{1992}]{Shu1992}
Shu, F. H., 1992, The physics of astrophysics, Volume II: Gas dynamics
 (\ufhref[webgreen]{http://adsabs.harvard.edu/abs/1992pavi.book.....S}{University Science Books})
 
\bibitem[\protect\citeauthoryear{Silich \& Tenorio-Tagle}{2013}]{Silich2013}
Silich, S., Tenorio-Tagle, G., 2013,
 \ufhref[webgreen]{http://dx.doi.org/10.1088/0004-637X/765/1/43}{ApJ},
 \ufhref[webgreen]{http://adsabs.harvard.edu/abs/2013ApJ...765...43S}{765, 43} 
  
 \bibitem[\protect\citeauthoryear{Skinner \& Ostriker}{2015}]{Skinner2015}
Skinner, M. A., \& Ostriker, E. C. 2015,
 \ufhref[webgreen]{http://dx.doi.org/10.1088/0004-637X/809/2/187}{ApJ}, 
 \ufhref[webgreen]{http://adsabs.harvard.edu/abs/2015ApJ...809..187S}{809, 187}

\bibitem[\protect\citeauthoryear{Spitzer} {1962}]{Spitzer1962}
Spitzer, L., Jr. 1962, Physics of Fully Ionized Gases,
 \ufhref[webgreen]{http://adsabs.harvard.edu/abs/1962pfig.book.....S}{(New York: Interscience)}
 

\bibitem[\protect\citeauthoryear{Str\"{o}mgren}{1939}]{Stromgren1939}
Str\"{o}mgren, B., 1939, 
 \ufhref[webgreen]{http://dx.doi.org/10.1086/144074}{ApJ},
 \ufhref[webgreen]{http://adsabs.harvard.edu/abs/1939ApJ....89..526S}{89, 526}
 
\bibitem[\protect\citeauthoryear{Taylor} {1950}]{Taylor1950}
Taylor, Geoffrey, 1950
 \ufhref[webgreen]{http://dx.doi.org/10.1098/rspa.1950.0049}{RSPSA}, 
 \ufhref[webgreen]{http://adsabs.harvard.edu/abs/1950RSPSA.201..159T}{201, 159}

\bibitem[\protect\citeauthoryear{Toro \etal} {1994}]{Toro1994}
Toro E. F., Spruce M., Speares W., 1994, 
 \ufhref[webgreen]{http://dx.doi.org/10.1007/BF01414629}{Shock Waves}, 
 \ufhref[webgreen]{http://dx.doi.org/10.1007/BF01414629}{4, 25}

\bibitem[\protect\citeauthoryear{Townsley \etal} {2006}]{Townsley2006}
Townsley, L. K., Broos, P. S., Feigelson, E. D., Brandl, B. R., Chu, Y., Gamire, G. P., \& Pavlov, G. G., 2006,
 \ufhref[webgreen]{http://dx.doi.org/10.1086/500532}{ApJ}, 
 \ufhref[webgreen]{http://adsabs.harvard.edu/abs/2006AJ....131.2140T}{131, 2140}
 
\bibitem[\protect\citeauthoryear{Wang \& Helfand} {1991}]{Wang1991}
Wang, Q., \& Helfand, D. J., 1991,
 \ufhref[webgreen]{http://dx.doi.org/10.1086/169840}{ApJ}, 
 \ufhref[webgreen]{http://adsabs.harvard.edu/abs/1991ApJ...370..541W}{370, 541}

\bibitem[\protect\citeauthoryear{Weaver \etal} {1977}]{Weaver1977}
Weaver, R., McCray, R., Castor, J., Shapiro, P., Moore, R., 1977,
 \ufhref[webgreen]{http://dx.doi.org/10.1086/155692}{ApJ},
 \ufhref[webgreen]{http://adsabs.harvard.edu/abs/1977ApJ...218..377W}{218, 377}

\bibitem[\protect\citeauthoryear{Wolfire \etal} {2003}]{Wolfire2003}
Wolfire, M. G., McKee C. F., Hollenbach, D., Tielens, A. G. G. M., 2003,
 \ufhref[webgreen]{http://dx.doi.org/10.1086/368016}{ApJ}, 
 \ufhref[webgreen]{http://adsabs.harvard.edu/abs/2003ApJ...587..278W}{587, 278}

\bibitem[\protect\citeauthoryear{Yadav \etal}{2016}]{Yadav2016}
Yadav, N., Mukherjee, D., Sharma, P., Nath, B. B. 2016,
 \ufhref[webgreen]{http://arxiv.org/abs/1603.00815v1}{arXiv:1603.00815v1}
 
\bibitem[\protect\citeauthoryear{Yeh \& Matzner} {2012}]{Yeh2012}
Yeh, S. C. C., \& Matzner, C. D. 2012, 
 \ufhref[webgreen]{http://dx.doi.org/10.1088/0004-637X/757/2/108}{ApJ},
 \ufhref[webgreen]{http://adsabs.harvard.edu/abs/2012ApJ...757..108Y}{757, 108}
\end{thebibliography}
\end{document}